\documentclass[twocolumn,tighten]{aastex6}

\usepackage{graphicx,amsmath,amssymb}
\newcommand{\kmps}{km\,s$^{-1}$}
\newcommand{\cgsbunit}{erg\,s$^{-1}$\,cm$^{-2}$\,arcsec$^{-2}$}

\begin{document}

\title{The [Ne {\sc iii}] Jet of DG Tau and Its Ionization Scenarios}

\author{Chun-Fan Liu\altaffilmark{1,2}, Hsien Shang\altaffilmark{1,2},
        Gregory J. Herczeg\altaffilmark{3}, and Frederick M. Walter\altaffilmark{4}}
\altaffiltext{1}{Institute of Astronomy and Astrophysics, Academia Sinica (ASIAA), 
Taipei 10617, Taiwan}
\altaffiltext{2}{Theoretical Institute for Advanced Research in Astrophysics
(TIARA), Academia Sinica, Taipei 10617, Taiwan}
\altaffiltext{3}{The Kavli Institute for Astronomy and Astrophysics, 
Peking University, Beijing 100871, China}
\altaffiltext{4}{Department of Physics and Astronomy, Stony Brook University, 
Stony Brook, NY 11794-3800, USA}

\begin{abstract}                                                      
     Forbidden neon emission from jets of low-mass young stars can be used 
to probe the underlying high-energy processes in these systems. We analyze 
spectra of the jet of DG Tau obtained with the Very Large 
Telescope/X-Shooter spectrograph in 2010. [Ne {\sc iii}] $\lambda3869$ 
is clearly detected in the innermost 3\arcsec\ microjet and the outer knot 
located at $\sim6\farcs5$. The velocity structure of the inner microjet can 
be decomposed into the low-velocity component (LVC) at $\sim -70$ \kmps\ and the 
high-velocity component (HVC) at $\sim -180$ \kmps. Based on the 
observed [Ne {\sc iii}] flux and its spatial extent, we suggest the origins
of the [Ne {\sc iii}] emission regions and their relation with known X-ray sources along the jet.
The flares from the hard X-ray source close to the star may be the main ionization source of 
the innermost microjet. The fainter soft X-ray source at $0\farcs2$ from the star
may provide sufficient heating to help to sustain the ionization fraction 
against the recombination in the flow. The outer knot may be reionized by shocks 
faster than 100 \kmps\ such that [Ne {\sc iii}] emission reappears and 
that the soft X-ray emission at $5\farcs5$ is produced. 
Velocity decomposition of the archival {\it Hubble Space Telescope}
spectra obtained in 1999 shows that the HVC had been faster, with a velocity centroid 
of $\sim -260$ \kmps. Such a decrease in velocity may potentially be 
explained by the expansion of the stellar magnetosphere, changing the truncation radius 
and thus the launching speed of the jet. The energy released by magnetic reconnections 
during relaxation of the transition can heat the gas up to several tens of megakelvin and 
provide the explanation for on-source keV X-ray flares that ionize the neon microjet.
\end{abstract}


\section{INTRODUCTION}\label{Intro}

Observations of kinematic structures and physical conditions of jets 
and winds from young stellar objects (YSOs) help to clarify how outflows
regulate the formation of young stars. Spectroscopic studies of
forbidden optical emission lines from classical T Tauri stars (TTSs) 
provide diagnostics of both kinematic and physical properties of
the jets and their driving region \citep{RB01}. 
Spatially resolved spectroscopy \citep{Bac00,Cof08} of
low-excitation, high-abundance line species such as [O {\sc i}], [S {\sc ii}],
and [N {\sc ii}] has been used to deduce physical conditions and kinematic
properties along the jet.

Forbidden lines of singly and doubly ionized neon
provide useful diagnostics of the physical conditions around the young stars 
\citep{GNI07,HG09} because of the high ionization potentials of neon.
The 12.81\micron\  [Ne {\sc ii}]  fine-structure
line has been detected toward low-mass YSOs associated with 
X-ray/ultraviolet photoevaporating disks \citep{PS09}, and in the terminal 
knots of jets and outflows from low-mass protostars \citep{vB09}. 
Photoionization by keV X-ray photons may be an important source
of [Ne {\sc ii}] excitation in disk atmospheres \citep{GNI07} and in jets and outflows \citep{SGLL}.
With a large critical density ($n_{\rm cr} \sim 10^7$ cm$^{-3}$ at
$T \sim 10^4$ K), the optical forbidden [Ne {\sc iii}] $\lambda3869$\AA\ transition
has been used as a marker for high-energy photons or for strong shocks 
in studies of high-mass star formation and active galactic nuclei.
Among the low-mass YSOs, detections of [Ne {\sc iii}] $\lambda3869$ close to 
the central star have been reported recently (on the scale of arcseconds),
mainly associated with their jets \citep{Liu14,Whelan14}. 

     DG Tau, located in the Taurus--Auriga molecular cloud at a distance 
of $\sim$140 pc \citep{KDH94,TLMR07}, is an actively accreting extreme T Tauri star 
with a spectrum consistent with a heavily veiled $\sim$K7 star \citep{HH14}.
The flat spectral energy distribution in the near- to mid-infrared \citep{KH95, Hm05}
suggests that DG Tau may represent a transition stage between the Class~I and Class~II YSOs.
The high optical veiling prevents an accurate determination of the extinction 
\citep{HEG95,GHBC98,WH04,HH14}. From modeling ultraviolet to optical continuum,
\cite{GCMH00} estimated a mass accretion rate of $\dot{M}_{\rm acc}\sim 5\times10^{-7}
M_{\odot}\,\rm{yr^{-1}}$ with a moderate extinction of $A_V \approx 1.6$ 
\citep[a value independently obtained through spectral typing by][]{HH14}. 
The bright [O {\sc i}] and [S {\sc ii}] emission 
is consistent with a strong outflow \citep{HEG95,WH04}; 
using $A_V \approx 1.6$, a mass-loss rate 
$\dot{M}_{\rm w}\gtrsim 1\times10^{-7}M_\odot\,{\rm yr}^{-1}$ was deduced 
using the [O {\sc i}] flux obtained by \cite{HEG95}.

     DG Tau drives an optically visible jet at a position angle of    
$\sim226\arcdeg$. Optical spectra have revealed it to be blueshifted \citep{SB93}. 
From the proper motion of the knots 5\arcsec\--10\arcsec\ from 
the star, an inclination angle of $37\fdg7\pm2\arcdeg$ with respect to the line of sight 
has been inferred \citep{EM98}. The inner 5\arcsec\ of the jet has been mapped using 
broad-band optical imaging with the {\it Hubble Space Telescope (HST)} \citep{WFPC2} and using 
adaptive optics (AO)-aided narrow-band imaging from the Canada-France-Hawai`i Telescope (CFHT) 
\citep{Dou00}. Both images showed a jet-like structure within 
2\arcsec\ of the source, and an emission peak at around 4\arcsec\ that 
resembles a bow shock. Using CFHT integral-field spectroscopy to investigate the kinematics 
at spatial scales smaller than 5\arcsec, \cite{LCD00} showed that the jet-like feature 
reaches its highest velocity of $\sim -350$ \kmps\ at $\sim 1\farcs3$ 
and drops to $\sim -280$ \kmps\ afterwards, before approaching 
$\sim -260$ \kmps\ at the 4\arcsec\ bow shock. Lower-velocity 
material ($\lesssim -100$ \kmps) appears to have a larger transverse spatial extent
and is concentrated closer to the star. The two distinct kinematic features
and their spatial and morphological differences have been 
further traced to within 2\arcsec\ with {\it HST}/Space Telescope Imaging Spectrograph (STIS) 
in optical forbidden lines \citep{Bac00,Bac02} and with the Subaru Telescope in the
near-infrared 1.644\micron\ [Fe {\sc ii}] line \citep{Pyo03}. 

The region of the innermost few arcseconds of the jet, often dubbed the ``microjet''
\citep{Solf97}, reveals information close to the jet-driving source.
{\it HST}/STIS spectra provide an opportunity to probe the kinematics of the DG Tau 
microjet at $\sim0\farcs1$ spatial resolution.
\citet{Bac00} constructed channel maps of H$\alpha$ and optical forbidden emission lines 
from these spectra. They binned spectra with velocities between 
$+70$ and $-420$ \kmps\ into four distinct velocity channels 
of approximately 125 \kmps\ width. The channels were 
designated as low-, medium-, high-, and very-high-velocity, 
respectively, indicating increasingly blueshifted radial velocities. 
Within 0\farcs7 of the star, the outflow has the form of a 
collimated jet. This is most evident in the high-velocity channel. A
bubble-like feature, most evident at intermediate velocities, 
is seen between 0\farcs4 and 1\farcs5 from the star. 
\cite{Maurri14} inferred physical conditions such as temperature and density along and
across the jet from line ratios.
The blueshifted jet was found to have a high density of $n_{\rm H} \sim 3\times10^6$ 
cm$^{\rm -3}$ and low electron fraction $x_e \sim 0.03$ and to reach an order-of-magnitude 
lower density and a high electron fraction of 0.7 at $\sim 0\farcs7$ and around $2\arcsec$ from the 
jet source. On the other hand, the physical conditions were found to alter little across the jet.
This spatial variations in physical conditions were interpreted as arising from ionization with
shocks weaker than $\sim 100$ \kmps.

DG Tau is an active and bright X-ray source.
It is the first known low-mass YSO to show spatially resolved X-ray emission along 
the jet axis, extending up to 5\arcsec\ from the star \citep{Gudel05,Gudel08}.
The extended X-rays have a luminosity of $\sim 10^{28}$ erg\,s$^{-1}$
and plasma temperature of $\sim 3$ MK \citep{Gudel08}.
It exhibits a proper motion of $\sim 0\farcs3\,{\rm yr}^{-1}$,
similar to other optical and infrared knots, leading to its identification as an
``X-ray jet'' \citep{Gudel12,DGTau_RadioJet}.
The on-source X-ray emission consists of a hard, flaring component with
$L_{\rm X} \sim 10^{30}$ erg\,s$^{-1}$ and $T_{\rm X} \sim 30$ MK 
and a soft, steadier component with $L_{\rm X}$ and $T_{\rm X}$ roughly
an order of magnitude lower than those of the hard component \citep{Gudel08}. 
Multi-year {\it Chandra} observations
show that the hard component is located at the stellar position and the
soft component is offset along the optical jet axis by 0\farcs2 \citep{SS08}.
The flaring hard component is attributed to the coronal emission common to YSOs;
its inferred column density of $N_{\rm H} \sim$ 2--3$\times10^{22}$ 
cm$^{-2}$ \citep[corresponding to $A_V = N_{\rm H}/2\times10^{21} \approx$ 10--15,][]{Vuong03} 
is much higher than the value ($A_V \approx 1.6$ or $N_{\rm H} \approx 3\times10^{21}$ 
cm$^{-2}$) obtained from optical-to-infrared photometry. The soft near-source component
has a spectrum similar to that of the extended X-ray jet and a lower column density $N_{\rm H}
\sim 1.1\times10^{21}$ cm$^{-2}$, corresponding to $A_V \approx 0.55$
\citep{Gudel08,Gudel12}. It may be associated with an inner part of the jet.
Spatially resolved {\it HST} spectra in the far-ultraviolet (FUV) C\,{\sc iv} doublet
show the visual proximity between the C\,{\sc iv} emission and 
the off-source soft X-ray component. This spatial correlation may suggest local 
heating up to $10^5$ K along the path of propagation of the jet \citep{Sch13L,Sch13}.
Understanding the roles of these multiple X-ray components associated with
the jet can help to elucidate the ionization and excitation of the jet upon launching and 
during propagation.

In this paper, we present spatially resolved spectroscopy of DG Tau's
jet observed with {\it HST}/STIS in 1999 and VLT/X-Shooter in 2010. In both spectra we 
identify double velocity components from the observed line profiles. 
In the X-Shooter spectra, [Ne {\sc iii}] $\lambda3869$ traces the jet up to 
8\arcsec\ from the star, with double velocity components within the inner 
3\arcsec. In Section \ref{obsana}, we present the observations and analysis of the two data sets. 
In Section \ref{results}, we describe the properties of the decomposed spectra, 
and the properties of the [Ne {\sc iii}] jet. Possible
origins of the [Ne {\sc iii}] emission in the DG Tau jet is discussed in Section 
\ref{discussion_neon} and the evolution of the velocity components
during the two observation epochs are discussed in Section \ref{discussion_vc}.
We summarize our findings in Section \ref{summary}.

\section{Observations and Analysis}\label{obsana}

\subsection{\textit{HST}/STIS Archival Data Analysis}\label{2dg}

\begin{figure*}
\includegraphics[height=\textwidth,angle=90]{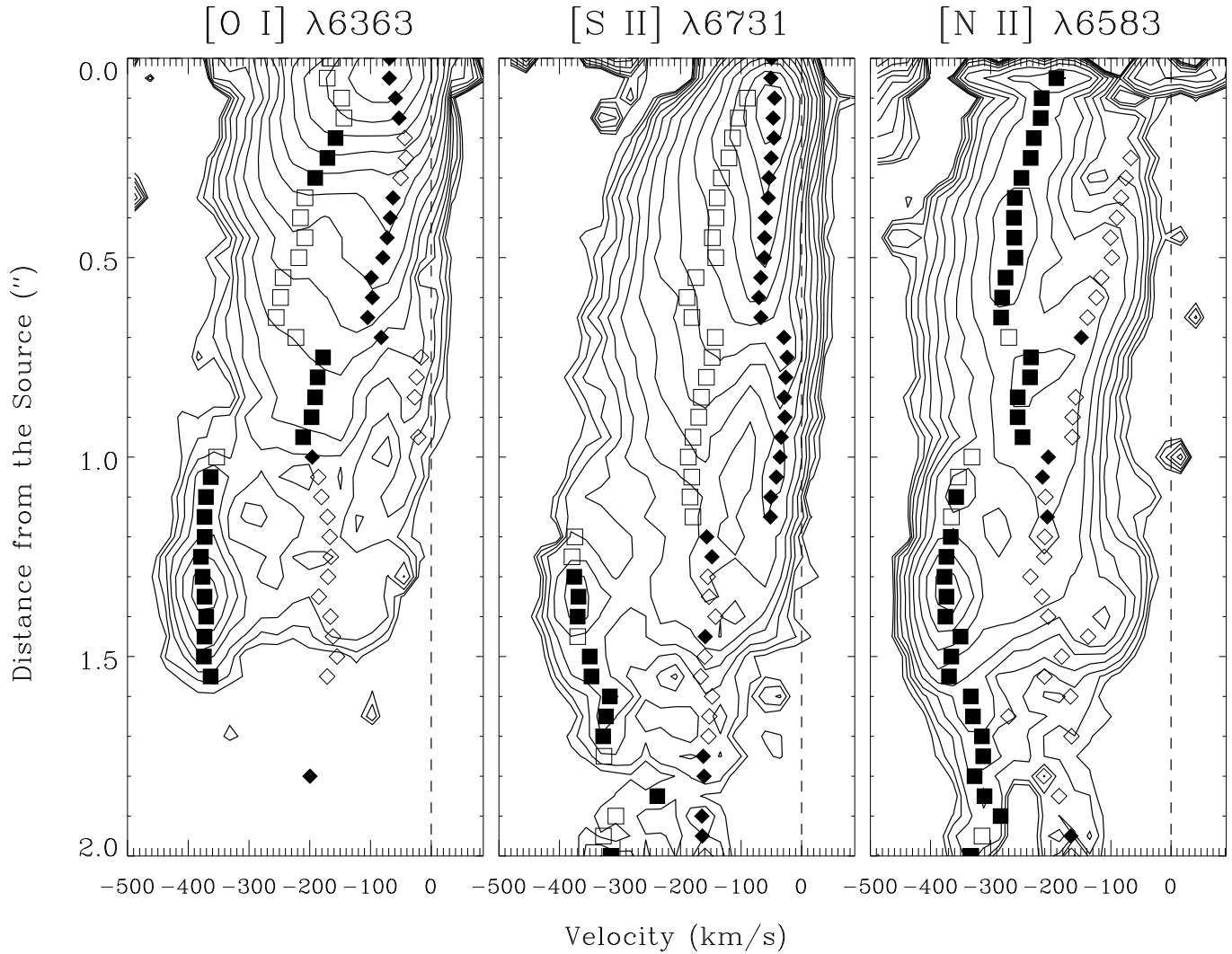}
\caption{The inner 2\arcsec\ PV diagrams of the DG Tau jet obtained by averaging 
        the {\it HST}/STIS spectra from seven slit positions perpendicular to the jet. 
        The emission lines of 
        [O {\sc i}] $\lambda6363$, [S {\sc ii}] $\lambda6731$, and [N {\sc ii}] $\lambda6583$, 
        are shown from left to right. The contours are 
        logarithmic with $\sqrt{2}\sigma$ increments, starting from $10\sigma$. 
        The $1\sigma$ uncertainties for [O {\sc i}], [S {\sc ii}], and [N {\sc ii}] 
        are $2.8\times10^{-16}$, $2.4\times10^{-16}$, and $2.2\times10^{-16}$ 
        \cgsbunit, respectively. The line profiles at each spatial position have
        been fitted and decomposed into two Gaussian profiles. The fitted velocity
        centroids of the decomposed profiles are shown by squares and diamonds
        for the high-velocity and low-velocity components, respectively.
        The filled and open symbols represent the stronger and weaker velocity
        components at each spatial position.}
\label{STIS_allPV}
\end{figure*}

\begin{figure*}
\epsscale{0.50} 
\plotone{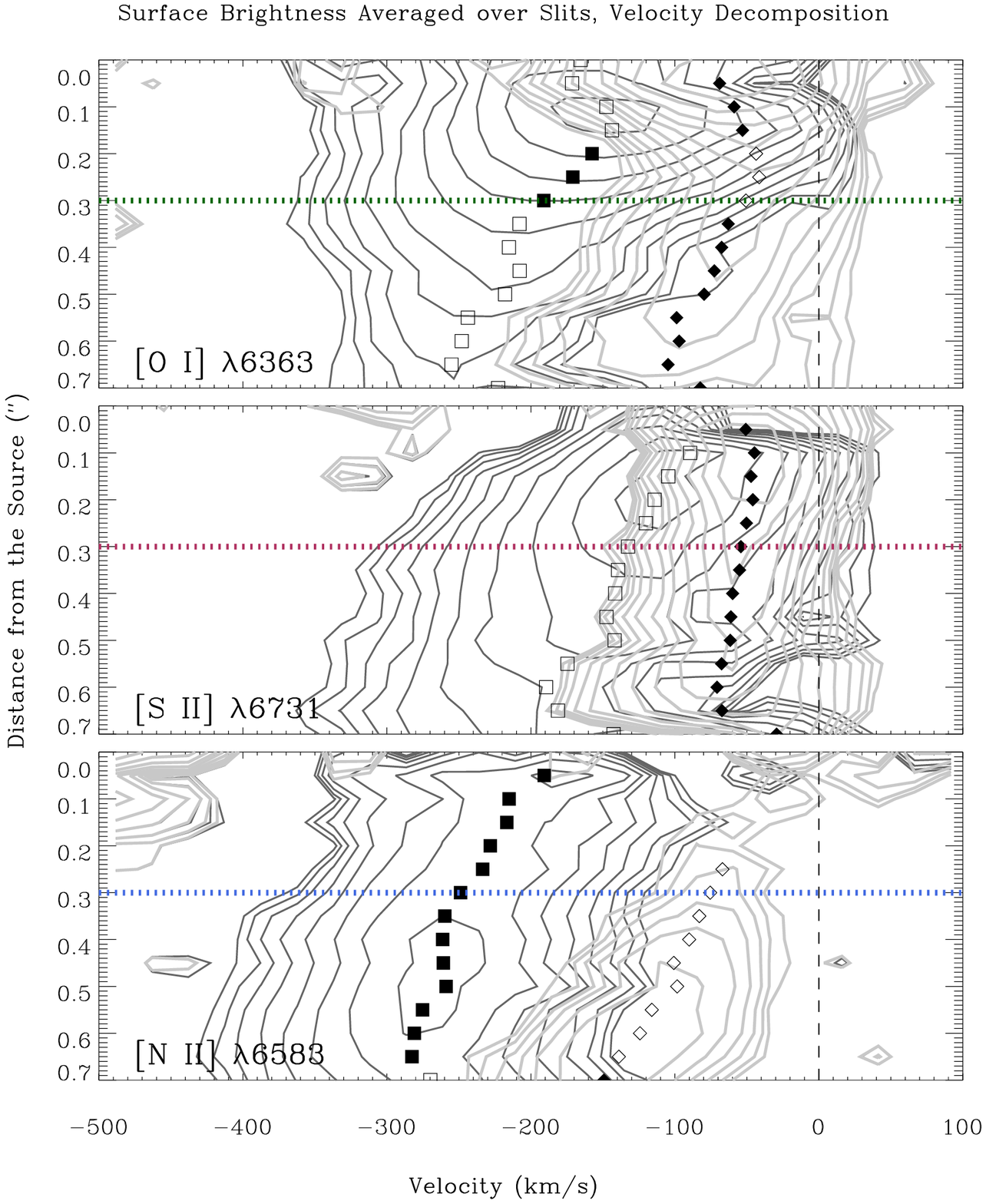}
\epsscale{0.45}
\plotone{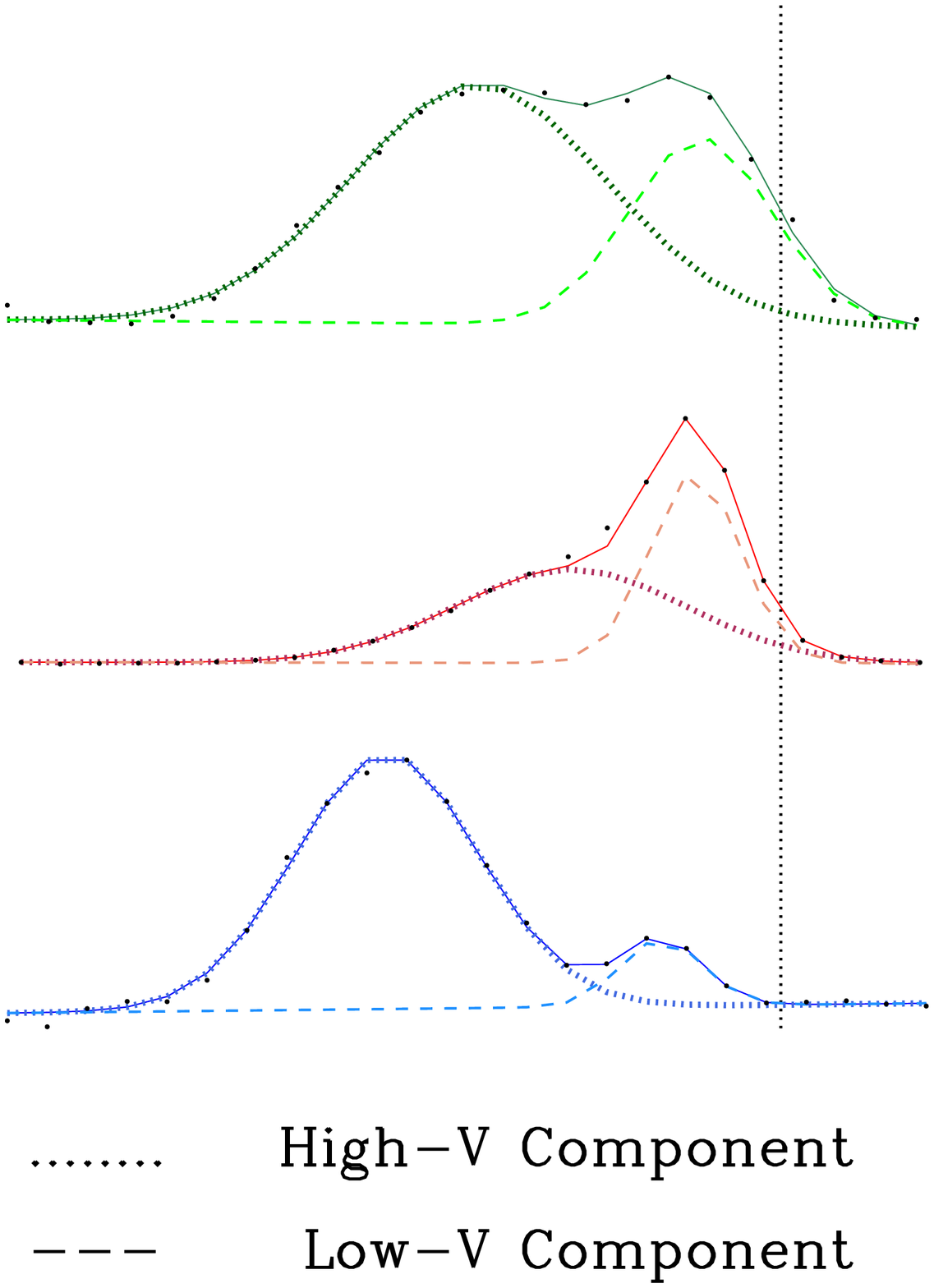}
\epsscale{1.0} 
\caption{The inner 0\farcs7 PV diagrams made by averaging 
         the {\it HST}/STIS spectra from seven slit positions, showing those of 
         [O {\sc i}] $\lambda6363$, [S {\sc ii}] $\lambda6731$, and [N {\sc ii}] $\lambda6583$, 
         from top to bottom. The contours are
         logarithmic with $\sqrt{2}\sigma$ increments, starting 
         from $10\sigma$. The $1\sigma$ uncertainties for [O {\sc i}], [S {\sc ii}], and [N {\sc ii}] 
         are $3.5\times10^{-16}$, $3.4\times10^{-16}$, and 
         $2.0\times10^{-16}$ \cgsbunit, respectively. The line profiles 
         are decomposed into a higher (black contours) and a lower (gray
         contours) velocity component with two Gaussians. The right panels
         show the observed line profiles (discrete points) and the two-Gaussian fits at a position 
         of 0\farcs3 to illustrate the concept of velocity decomposition.
         The overall (summed) line profiles are shown by thin solid lines.
         The fitted HVC profiles are shown by thick dashed Gaussians and the fitted LVC 
         profile are shown edit2{by} thin dashed Gaussians.
         The centroids from two-Gaussian fitting are plotted with
         squares (the lower velocity Gaussian) and diamonds (the higher 
         velocity Gaussian) are plotted on top of the contours in the left panels.
         The filled symbols show the Gaussian with 
         higher intensity and open symbols the Gaussian with lower intensity.
         The results indicate the different nature of the three lines, 
         with the [O {\sc i}] and [S {\sc ii}] lines dominating at low velocity ($< 200$ 
         \kmps) before 0\farcs7, and the [N {\sc ii}] line peaking at high velocity 
         ($\sim 260$ \kmps) throughout the flow.}
\label{STIS_avgPV}
\end{figure*}

We downloaded pipeline-processed DG Tau spectra of the {\it HST} Cycle 7 
observations from the Mikulski Archive for Space Telescopes (MAST). 
The observations were taken on 1999 January 14 (GO 7311, 
PI: R. Mundt) with the STIS.
The observational settings have been described by \cite{Bac00}
and \cite{Maurri14}, so we summarize only the essential properties here.
The $\mathtt{52\times0.1}$ slit and {\tt G750M} grating 
centered at {6581\AA} were used to cover six bright optical forbidden 
emission lines and H$\alpha$.
Seven slit positions parallel to the axis of the DG Tau jet (P.A.$=226\arcdeg$) 
were observed, each offset by 0\farcs07 in the transverse direction to cover a total span 
of 0\farcs52 across the jet. Each exposure yielded a 
two-dimensional spectral image covering 52\arcsec\
(0\farcs05 pixel$^{-1}$ or $\sim 0\farcs1$ FWHM) perpendicular to the jet,
and 6295--6867 \AA\ along the dispersion axis 
(0.554\AA\ pixel$^{-1}$, corresponding to $\sim25$ km\,s$^{-1}$),
with a velocity resolution of $\sim 65$ \kmps.

The pipeline-processed spectra from MAST suffice for data
analysis. Hot pixels were present primarily in the line-free regions.
Bad pixels that affect a few rows of the spectra between 
1\arcsec\ and 2\arcsec\, were flagged by inspection.
Further reductions primarily consisted of removing the stellar 
continuum and the contribution from the reflecting nebula. Each spectral 
image was first divided into three sub-images containing lines of 
[O {\sc i}], H$\alpha+$[N {\sc ii}], and [S {\sc ii}], respectively. 
On each sub-image, the rows containing the blueshifted jet extending 
up to 2\arcsec\ from the star were examined to remove the semi-periodic baseline 
undulations resulting from wavelength rectification of the undersampled 
star. To each row, we fit and subtract the baseline with a 
Legendre polynomial up to the tenth order, depending on the distance to 
the star. The tasks were performed within IRAF/STSDAS {\tt GFIT1D} using 
the amoeba $\chi^2$ minimization.

We extracted position--velocity (PV) diagrams for [O {\sc i}] 
$\lambda\lambda6300/6363$, [N {\sc ii}] $\lambda\lambda6548/6583$, and 
[S {\sc ii}] $\lambda\lambda6717/6731$. We converted wavelengths to
radial velocities, assuming the systemic velocity of DG Tau to be 
$+16.5$ km\,s$^{-1}$ \citep{Bac00}. PV diagrams were extracted
from each of the seven slit positions, and also 
from transversely averaged spectral images to cover the emission across the jet
and raise the signal-to-noise ratio. PV diagrams of the
jet up to $-500$ km\,s$^{-1}$ can be extracted, except for [O {\sc i}] 
$\lambda6300$, which is cut off at $\sim -230$ km\,s$^{-1}$ by the end of the detector.
We therefore used [O {\sc i}] $\lambda6363$
for kinematic studies of the [O {\sc i}] emission whenever applicable.
Figure \ref{STIS_allPV} shows the baseline-subtracted PV diagrams for the 
innermost 2\arcsec\ of the jet, averaged over the seven slits across 0\farcs52 
perpendicular to the jet axis. [O~{\sc i}] $\lambda6363$, [S {\sc ii}]
$\lambda6731$, and [N {\sc ii}] $\lambda6583$ are shown.

Astrophysical emission lines are generally not Gaussian, but, where 
appropriate, single Gaussian fits do provide robust estimates of the bulk 
velocity (the line centroid) and the line broadening.
Visual inspection of the PV diagrams shows that single Gaussians often do not
suffice to describe the observed line profiles, so we decomposed the spectra into
different velocity components by fitting two independent Gaussian profiles every 0\farcs05.
These line profiles can appear double-peaked, flat-topped, or skewed, depending on
the relative fluxes of gases at different velocities.
This can be visualized through the V-shaped isocontours in the PV diagrams for the 
innermost 1\arcsec\ of the jet (Figure \ref{STIS_allPV}), which is suggestive of 
double velocity components. Indeed, it has long been known that double velocity 
components exist in the DG Tau jet \citep{SB93,Pyo03}.
While the fitting and the decomposition are a mathematical exercise, the component centroids
and widths constrain the true physical velocity distributions of the gas.
Where two distinct Gaussian components are mathematically required, we interpret
the two components as distinct regions at different bulk velocities. The variations of these
principal velocity components with position along the jet can be either continuous or
abrupt depending on the propagation history of the jet. The relative fluxes of these principal
velocity components may also change along the flow. These variations provide insights into
the ejection history and possible variations of properties close to the
jet launching region.

We used the IRAF/STSDAS {\tt NGAUSSFIT} task to interactively give initial guesses 
for the Gaussian parameters and then recursively minimized $\chi^2$ with 
the amoeba algorithm. The concept of velocity decomposition and the 
resulting decomposed PV diagrams of the DG Tau microjet within 
0\farcs7 from the star are shown in Figure \ref{STIS_avgPV}. [O {\sc i}]
$\lambda6363$, [S {\sc ii}] $\lambda6731$, and [N {\sc ii}] $\lambda6583$
are shown, along with their line profiles at 0\farcs3 to show the representative
decompositions using two-Gaussian fitting.

\subsection{VLT/X-Shooter Optical Spectra}

\begin{figure*}
\plotone{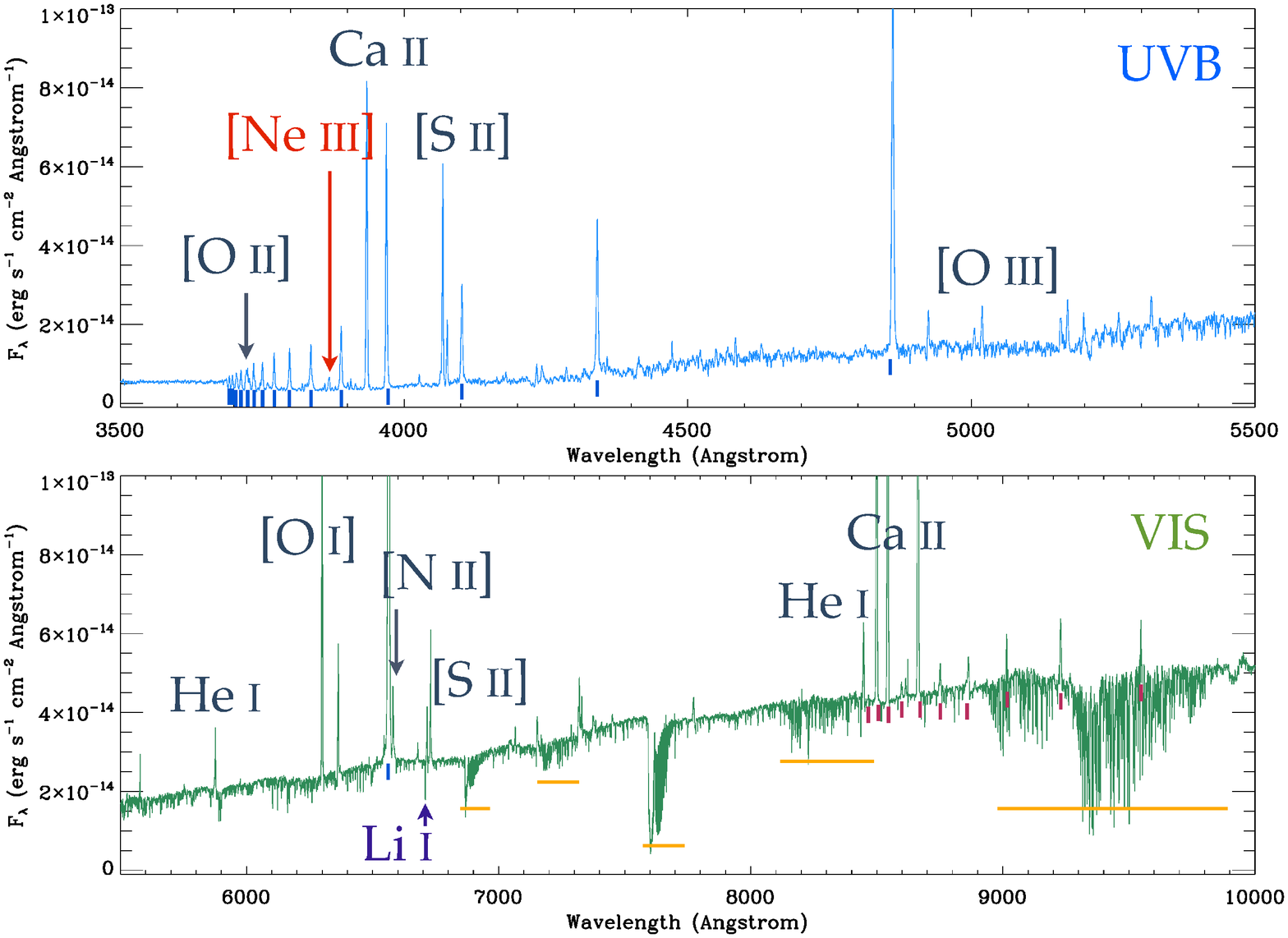}
\caption{One-dimensional spectra of DG Tau from the UVB (top)
         and VIS (bottom) arms by summing over $\pm3\arcsec$ from the 
         stellar position. The blue vertical bars show the identifications 
         of the Balmer series and the red vertical bars show the Paschen series.
         The spectra were not corrected for telluric absorption;  regions
         dominated by telluric absorption are marked with yellow horizontal bars.
         Other bright permitted lines (He {\sc i} and Ca {\sc ii}) and 
         forbidden lines ([O {\sc i}], [O {\sc ii}], [O {\sc iii}], 
         [S {\sc ii}], [N {\sc ii}]) are labeled. The location of 
         [Ne {\sc iii}] $\lambda3869$ is shown in red in the top panel.}
\label{dgtau_1d_uvb+vis}
\end{figure*}

New long-slit spectra of DG Tau and its jets were obtained with the 
X-Shooter spectrograph on the Very Large Telescope (VLT) on 2010 January 19
(084.C-1095(A), PI: G. Herczeg). A spatial coverage of 21\arcsec\ was obtained 
by using an 11\arcsec\ long slit in the nodding mode with two nodding 
positions separated by 10\arcsec\ at the position angle of 227\degr, roughly 
along the jet axis. An {\tt ABBA} nodding pattern was used.
The three spectral arms simultaneously cover the
UVB (300--550 nm), VIS (550--1000 nm), and NIR (1000--2500 nm) ranges, with
pixel scales of 0\farcs2\,pixel$^{-1}$ in the spatial direction and 0.02 
nm\,pixel$^{-1}$ in the dispersion direction ($\sim 15$, $8$, and $4$ 
km\,s$^{-1}$ for the three arms, respectively).
Exposure times for each nodding position were 240, 250, and 15 s for the
three arms, respectively. Slit widths of 1\farcs3, 1\farcs2, and 1\farcs2
were used for the three arms, respectively, resulting in a spectral resolution
of $R\sim5000$ ($\Delta v \sim 60$ km\,s$^{-1}$).
The seeing was $\sim1\arcsec$ throughout the 
observation and within the observed spectral range.

We confine our reduction and analysis to the UVB and VIS arms.
The spectra were reduced with the X-Shooter pipeline version 2.4.0 using
the {\it EsoRex} version 3.10.2. 
The reduction cascade of the stare mode was adopted instead of the nodding 
mode, since both nodding positions were used as the source frame. 
In the pipeline, bias and dark subtraction, flat-fielding, wavelength calibration, sky 
subtraction, and flux calibration were performed and two-dimensional spectra 
were obtained after concatenating adjacent orders. Two exposures of the 
same nodding positions were combined during the reduction cascade, and the 
combined data of the two nodding positions were connected by matching the 
spectrally summed spatial profiles along the jet axis. The resulting spectra 
contain the first $\sim 14\farcs2$ of the blueshifted jet and $\sim 6\farcs8$ 
of the redshifted counterjet. Median filters were used to remove the 
remaining hot and bad pixels to produce the final flux-calibrated and 
sky-subtracted two-dimensional spectral images. Figure \ref{dgtau_1d_uvb+vis} 
shows the reduced one-dimensional spectrum of DG Tau and its microjet,
extracted within $\pm3\arcsec$ of the jet axis.
Li {\sc i} $\lambda6707$ is detected at 6708.8 \AA, at a 
velocity shift of $+43.1$ \kmps, which we adopted as the 
systemic velocity offset for this data set.

To extract line properties, we removed the continuum emission in the spectra by 
fitting the line-free regions with polynomials. The combined spectra were first 
divided into segments of spectral ranges. Most of the segments follow spectral 
orders, except for the regions between 700--800 nm and between 900--1000 nm in 
the VIS arm, in which two orders were combined to cover telluric absorption 
features. In each segment, line-free regions were selected by consecutive 
moving median filters of increasing box sizes and fine-tuned by cross-checking 
with the emission line database and by inspection. The fitting and subtraction 
were performed at each row along the slit; the order of the polynomial was varied 
from 3 to 7 depending on the emission features in the individual row. 

PV diagrams were created from the continuum-subtracted two-dimensional spectra.
Two-Gaussian velocity decomposition was performed as described in Section \ref{2dg}.
Selected PV diagrams of emission lines from the UVB and VIS arms 
are shown in Figures \ref{PV_UVB} and \ref{PV_VIS}, 
respectively. The fitted velocity centroids are overlaid on the PV diagrams
and are further described in Section \ref{VLTobs}.

\section{RESULTS}\label{results}

\subsection{Kinematics of the Microjet from Archival {\it HST}/STIS Spectra} \label{HSTobs}

     Using the archival {\it HST}/STIS spectra taken in 1999, we obtained 
PV diagrams of the blueshifted microjet of DG Tau covering positions within 
2\arcsec\ of the star (Figure \ref{STIS_allPV}).
In this figure, the velocity centroids measured by two-Gaussian decomposition are 
overlaid on the PV diagrams, with the squares representing the centroids of the 
high-velocity component (HVC) and the diamonds representing those of the low-velocity component 
(LVC). At each spatial position, the stronger component is rendered with the filled 
symbol and the weaker component with the open symbol.
Where a single Gaussian fit suffices (as at distances 0\farcs75--0\farcs95 
and 1\farcs0--1\farcs6 in the [O {\sc i}] and [N {\sc ii}] 
PV plots), the filled square symbols represent the sole velocity. 

Three kinematic regions are evident within the innermost 2\arcsec.
The first region, from the star out to $\sim 0\farcs7$,
is characterized by a smooth change in velocity centroids.
This we regarded as the innermost undisturbed microjet. The second region starts from
$\sim 0\farcs7$, where a velocity discontinuity is seen in both the HVC and LVC.
Spatially, this coincides with the A2 region identified 
in the literature \citep{Bac00,Maurri14}. Within this region, the
HVC and LVC centroids of [O {\sc i}] and [S {\sc ii}] emission decrease by $\sim 100$ and $\sim50$
\kmps, respectively; on the other hand, the two components of [N {\sc ii}] seems to merge.
Another discontinuity from $\sim -200$ \kmps\ to $\sim -350$ \kmps\ marks the 
start of the third region. In the [O {\sc i}] and [N {\sc ii}] lines, 
the velocity jump occurs at $\sim 1\arcsec$; this change appears to occur about 
0\farcs2 further out in [S {\sc ii}]. For all the lines, the emission peaks 
at $\sim1\farcs35$ with a velocity centroid of $\sim -360$ \kmps 
and then becomes fainter and less ordered downstream.
This region coincides with A1 \citep{Bac00,Maurri14}.
Table \ref{tab_stis} shows the kinematic properties of 
[O {\sc ii}], [S {\sc ii}], and [N {\sc ii}] in the three regions within 2\arcsec\ 
of the star, obtained from the Gaussian decomposition analysis. 
For each region, the median value and standard deviation from within the spatial
extent of the region are shown.

Figure \ref{STIS_avgPV} shows the velocity decompositions of the transversely 
averaged PV diagrams of the jet. The innermost 0\farcs7 region is 
shown, corresponding to the undisturbed innermost microjet. In the left panels we 
show PV diagrams of the decomposed velocity components as two sets of contours:
the dark and light contours represent the HVC and the LVC, respectively.
The HVC is defined as the remainder after 
subtracting the LVC Gaussian fits from the spatial profile, and 
vice versa. In the right panels we show the actual line profiles 
and their fits at the representative 
position of 0\farcs3 to both illustrate the concept of velocity decomposition
and show the differences between the profiles of the three lines. 

     The overall line profiles of the three species differ from each other, as shown 
in the representative line profiles in the right panels of Figure \ref{STIS_avgPV}. 
The [O {\sc i}] LVC dominates close to the star and the profiles
appear flat-topped. The HVC becomes discernable beyond 0\farcs2 and forms a second peak 
blended with the LVC peak. Although blended, the peaks are typically separated by
$\sim100$ \kmps, which is larger than the velocity resolution of $\sim 65$ \kmps.
The [S {\sc ii}] LVC is always stronger than its HVC, and 
the line shape is typically a clean LVC peak at $\sim -60$ \kmps\ 
with a blue ``shoulder'' that extends from $\sim -140$ \kmps\ to 
$\sim -300$ \kmps\ and skews the overall profile. In [N {\sc ii}] the dominant emission is
from the HVC, but the LVC can be identified as a weak but distinct peak
beyond 0\farcs2 from the star. In all three line species, two-Gaussian fitting
would be necessary to fully describe the line profiles.

     The velocity structures along the jet differ among the three 
line species. Overall, the velocity centroids gradually become bluer as the 
distance increases from the star. The [N {\sc ii}] HVC changes from $-200$ 
to $-300$ \kmps\ within 0\farcs7, and its LVC is systematically offset by $\sim 150$~\kmps.
[O {\sc i}] behaves similarly, but is systematically
$\sim 40$ \kmps\ slower than the [N {\sc ii}] components. [S {\sc ii}] has
the slowest HVC, ranging from $-90$ to $-200$ \kmps, some $100$
\kmps\ slower than the [N {\sc ii}] HVC; its LVC is relatively steady along
the jet between $-40$ and $-70$ \kmps.

Comparison of the kinematic properties of the HVCs suggests
that these lines trace the same bulk flow but are 
excited at different regions within the flow. 
All the HVCs have broad line widths, 
in the range of 130--200 \kmps. The [N {\sc ii}] HVC line width is
fairly stable at $\sim 150$ \kmps. The [S {\sc ii}] line width is 
maximal ($\sim 200$ \kmps) at 0\farcs5, while [O {\sc i}] has its largest 
line width at 0\farcs3.

     The properties of the LVCs of [O {\sc i}] and [S {\sc ii}] are more
similar to each other than to that of [N {\sc ii}]. In the [O {\sc i}] and 
[S {\sc ii}] lines, the LVCs can be traced close to the vicinity of the star. 
The [N {\sc ii}] LVC, however, is not visible within the first 0\farcs2 of the jet.
The [O {\sc i}] and [S {\sc ii}] LVCs show a fairly constant velocity with 
decreasing brightness up to 0\farcs7. The brightness and velocity 
centroids of the [N {\sc ii}] LVC both increase along the jet until this 
component merges with its HVC at $\sim 0\farcs7$. For the three species, 
the LVC velocity centroids range from $-50$ to $-100$ \kmps. 
The typical [S {\sc ii}]$\lambda6731$ and [O {\sc i}]$\lambda6300$ LVC velocity centroids,
near $-50$ \kmps, are consistent with the values reported by \cite{HEG95}.
For all three line species the line widths of the LVCs are narrower than 
those of the HVCs, ranging from $\sim 70$ 
\kmps\ for [S {\sc ii}] to $\sim 120$ \kmps\ for [O {\sc i}]. 
The line profiles of the LVCs appear symmetric about the velocity centroids. 

\begin{table}[ht]
\begin{deluxetable*}{lccc}
 \tablewidth{0pt}
 \tablecaption{The Inner 2\arcsec\ Kinematic Properties from the 1999 {\it HST}/STIS Spectra\label{tab_stis}}
 \tablehead{
  \colhead{} & \colhead{[O {\sc i}] $\lambda6363$} &
               \colhead{[S {\sc ii}] $\lambda6731$} &
               \colhead{[N {\sc ii}] $\lambda6583$} \\
  \colhead{Component (Region)} & \multicolumn{3}{c}{median $\pm$ std dev. (\kmps)}
 }
\startdata
LVC ($d<0\farcs7$) &          $-67.47\pm22.6$ & $-59.86\pm8.45$ & $-98.38\pm23.5$ \\
HVC ($d<0\farcs7$) &          $-207.9\pm36.8$ & $-141.5\pm27.5$ & $-259.8\pm21.5$ \\
LVC ($0\farcs7<d<1\farcs1$) & $-27.15\pm83.5$ & $-27.64\pm3.61$ & \nodata \\ 
HVC ($0\farcs7<d<1\farcs1$) & $-210.9\pm92.1$ & $-165.6\pm12.1$ & $-255.1\pm59.1$ \\
LVC ($d>1\farcs1$) &          $-170.0\pm49.5$ & $-148.1\pm48.5$ & $-204.2\pm34.8$ \\
HVC ($d>1\farcs1$) &          $-373.3\pm56.3$ & $-350.1\pm83.0$ & $-350.1\pm26.6$
\enddata
\end{deluxetable*}
\end{table}

\subsection{Kinematics of the Microjet from VLT/X-Shooter Spectra}\label{VLTobs}

\begin{figure*}
\includegraphics[width=0.2\textwidth]{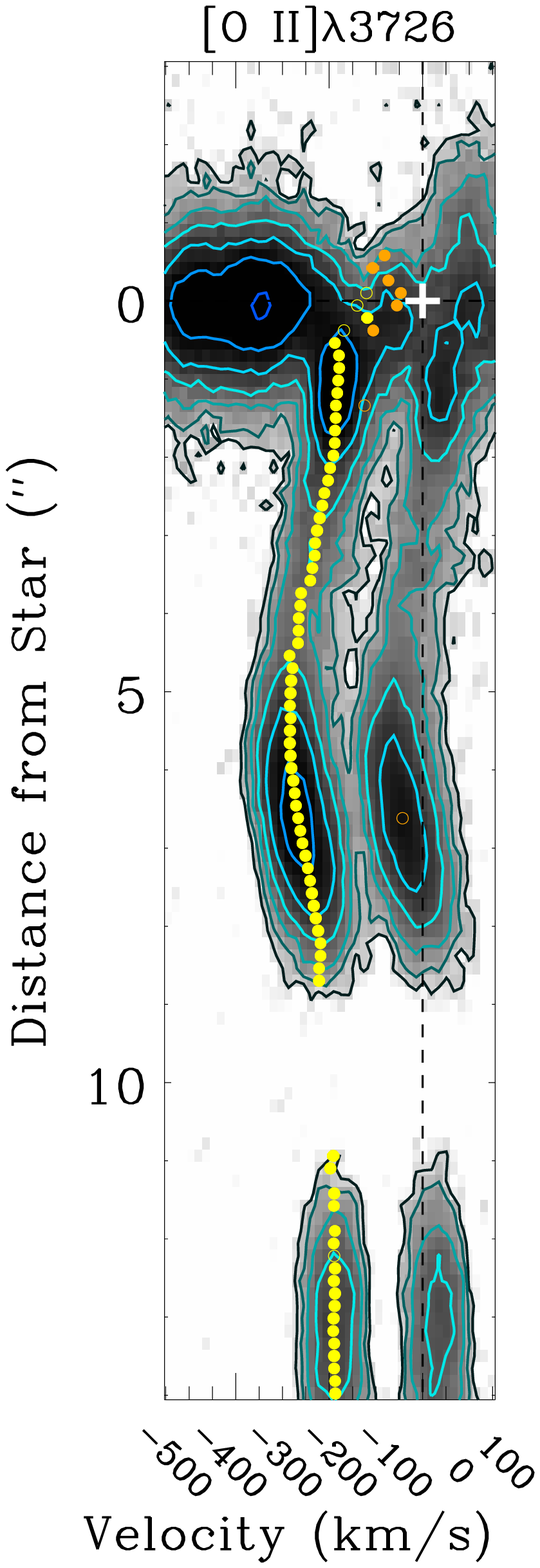}  
\hspace*{-0.2cm}
\includegraphics[width=0.2\textwidth]{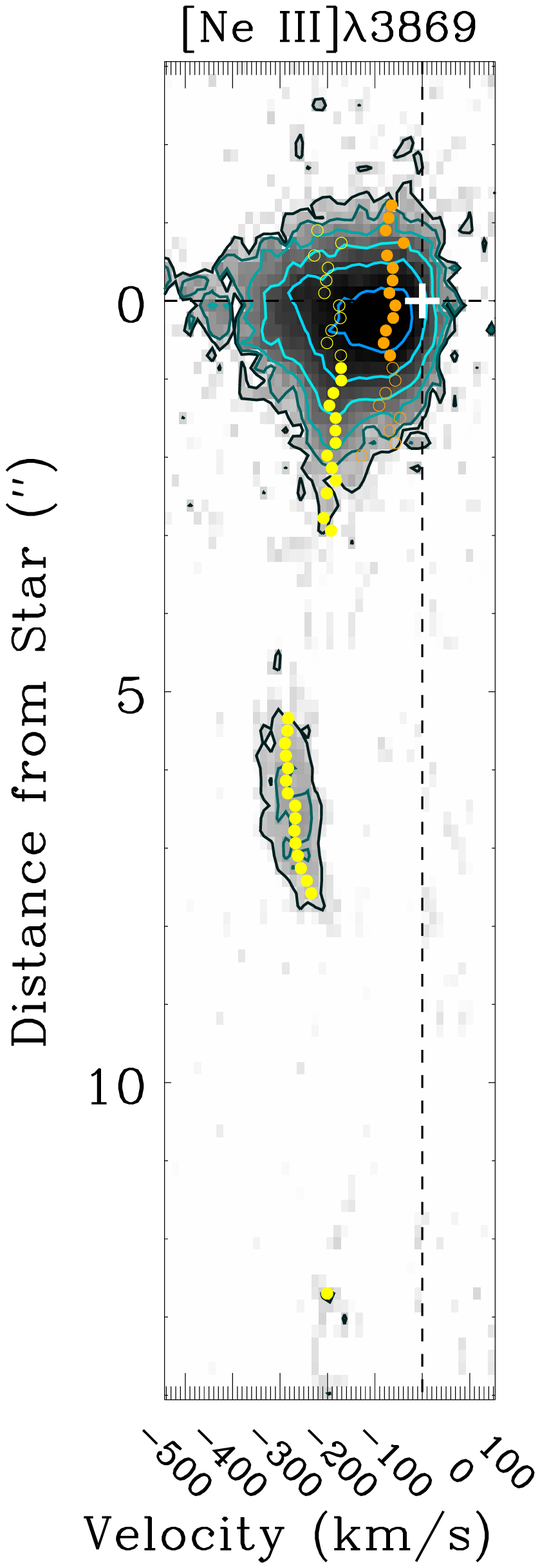}
\hspace*{-0.2cm}
\includegraphics[width=0.2\textwidth]{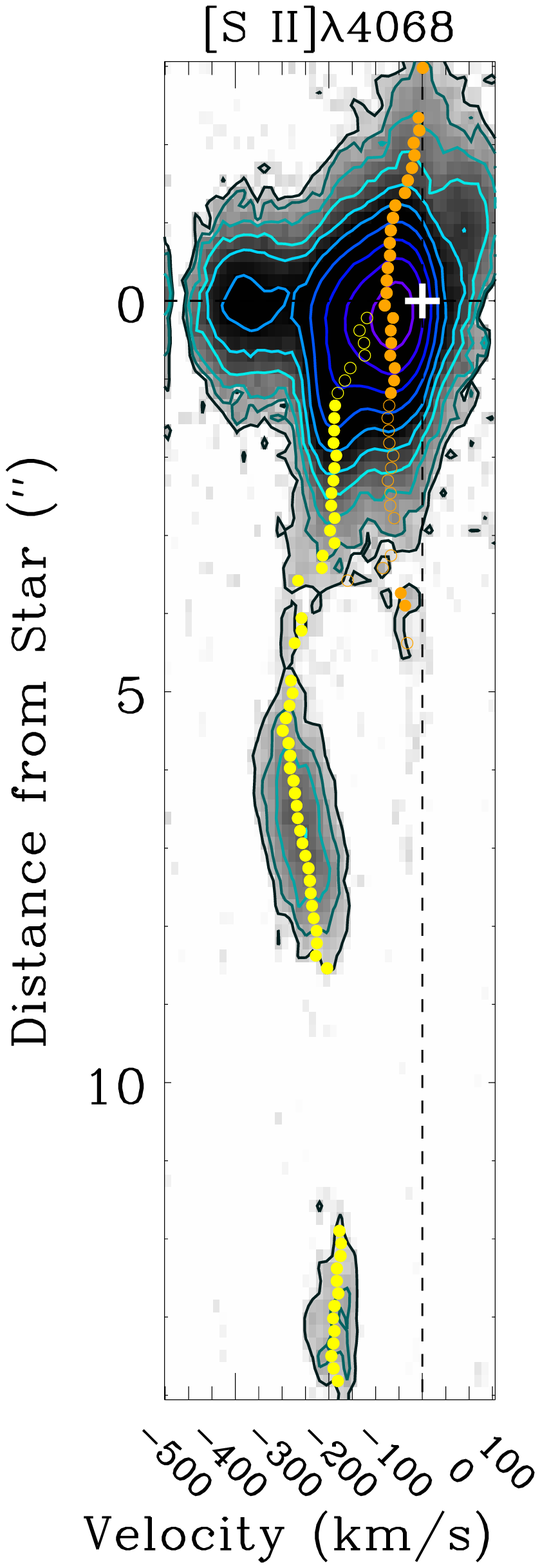}  
\hspace*{-0.2cm}
\includegraphics[width=0.2\textwidth]{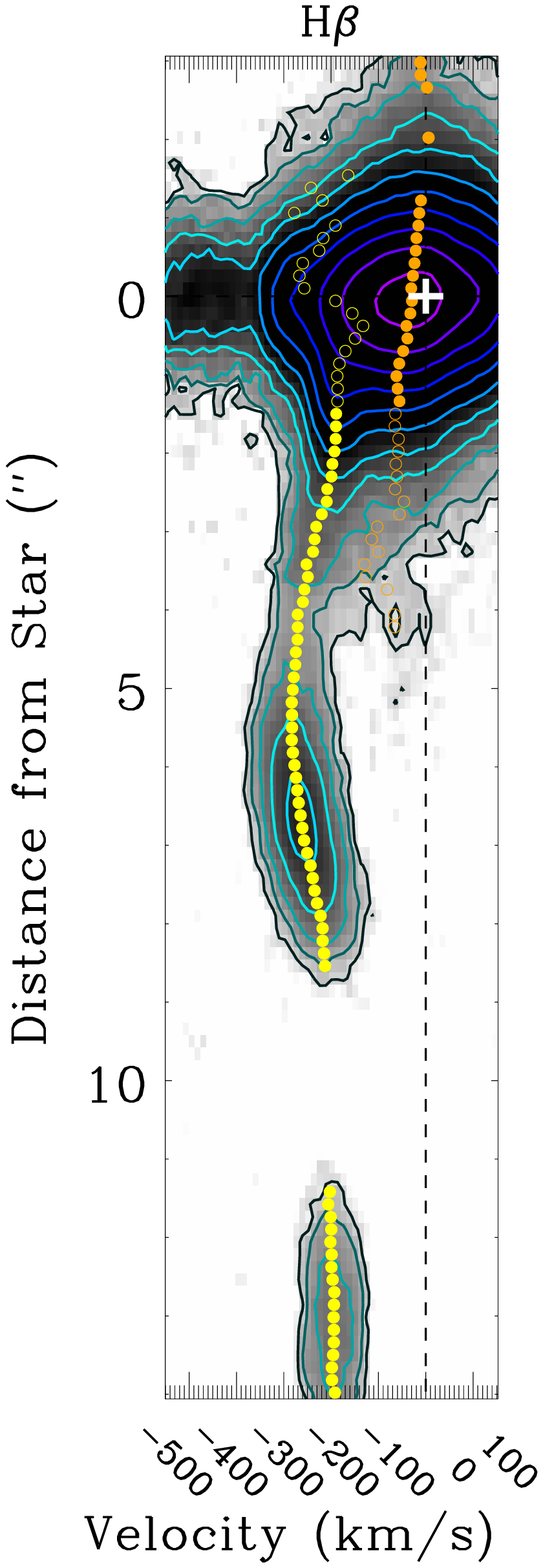} 
\hspace*{-0.2cm}
\includegraphics[width=0.2\textwidth]{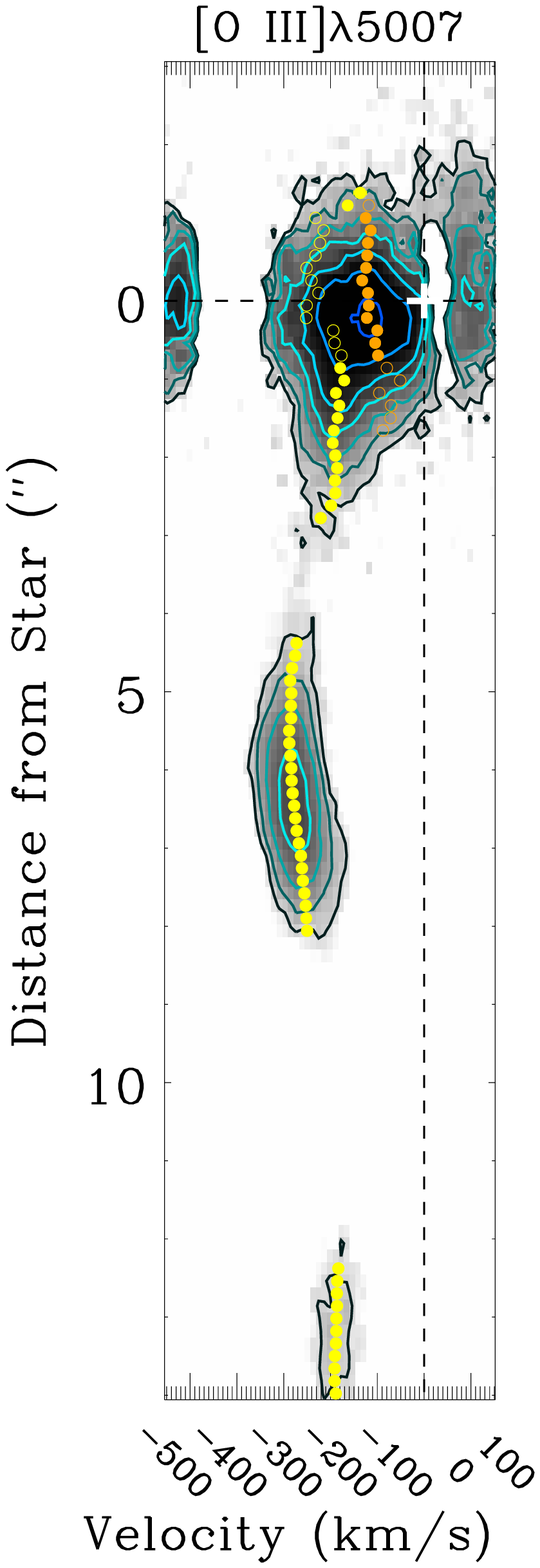} 
\hspace*{-0.2cm}
\caption{Selected PV diagrams of bright emission lines from 
         the VLT/X-Shooter UVB spectra, showing [O {\sc ii}] $\lambda3726$,
         [Ne {\sc iii}] $\lambda3869$, [S {\sc ii}] $\lambda4068$, 
         H$\beta$, and [O {\sc iii}] $\lambda5007$.
         The line profiles have been decomposed into high-velocity (yellow)
         and low-velocity (orange) components. At each position, the fitted 
         velocity centroids of the components are overlaid on the PV diagrams,
         with the stronger component shown by filled circles and the weaker component
         by open circles. In all the PV diagrams, the contours start from $3\sigma$ 
         ($1\sigma=3\times10^{-18}$ erg\,s$^{-1}$\,cm$^{-2}$\,\AA$^{-1}$)
         and increase by factors of 2.}
\label{PV_UVB}
\end{figure*}

\begin{figure*}
\includegraphics[width=0.2\textwidth]{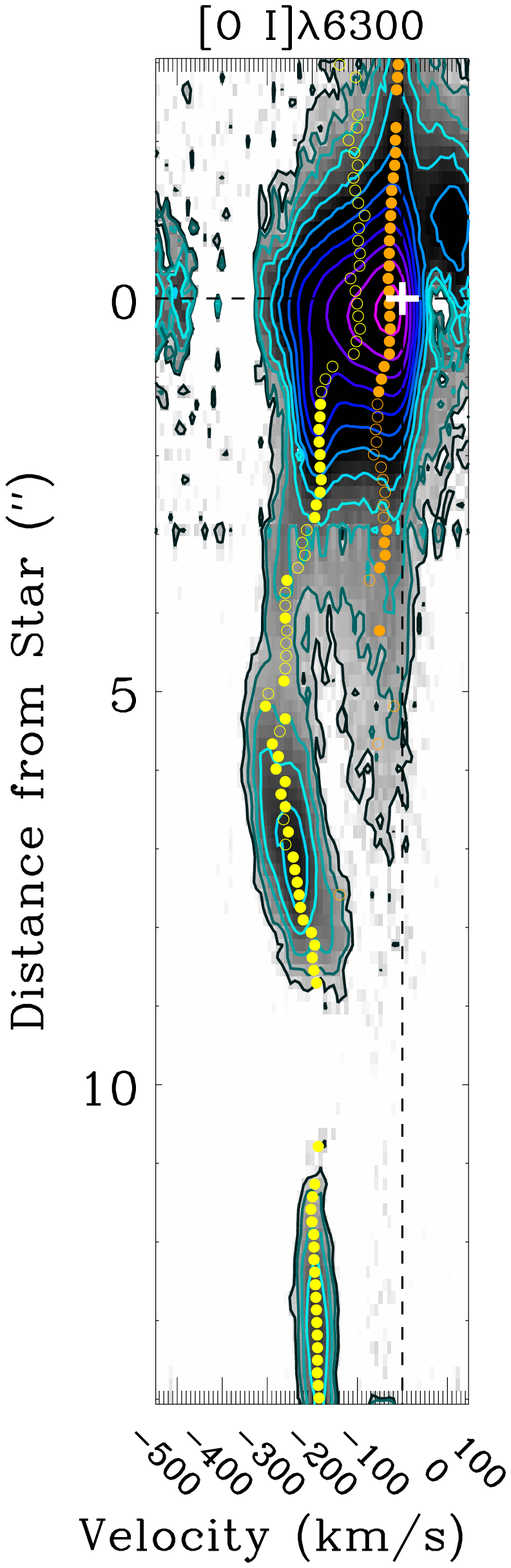}  
\hspace*{-0.2cm}
\includegraphics[width=0.2\textwidth]{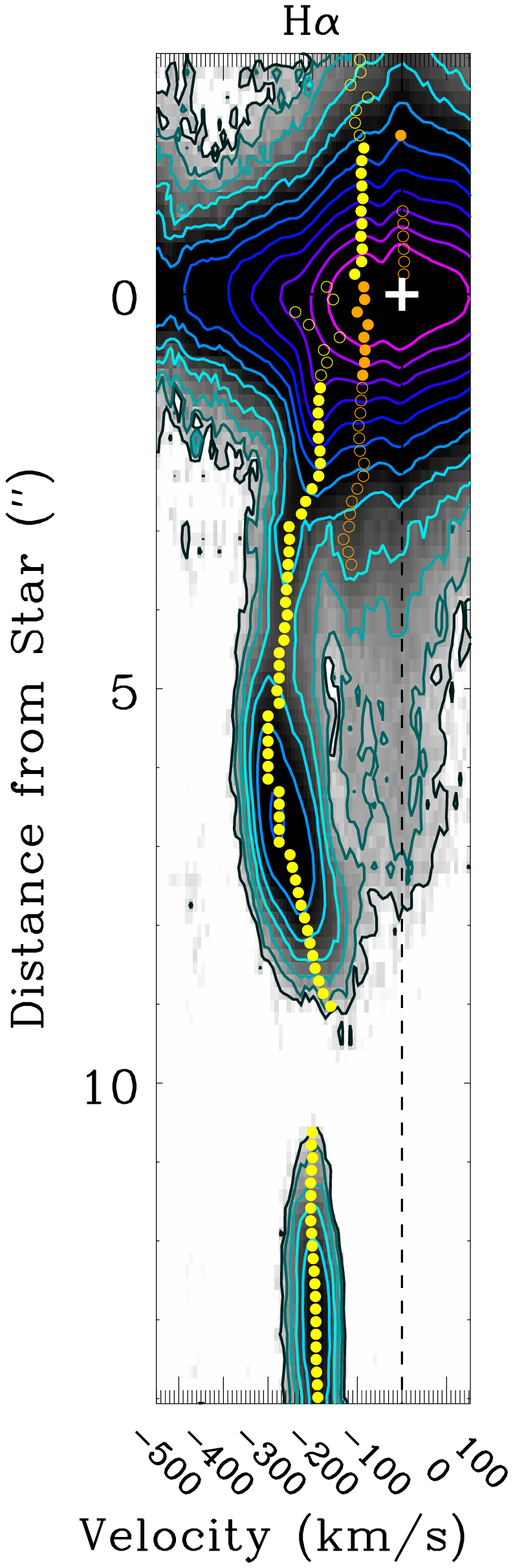}  
\hspace*{-0.2cm}
\includegraphics[width=0.2\textwidth]{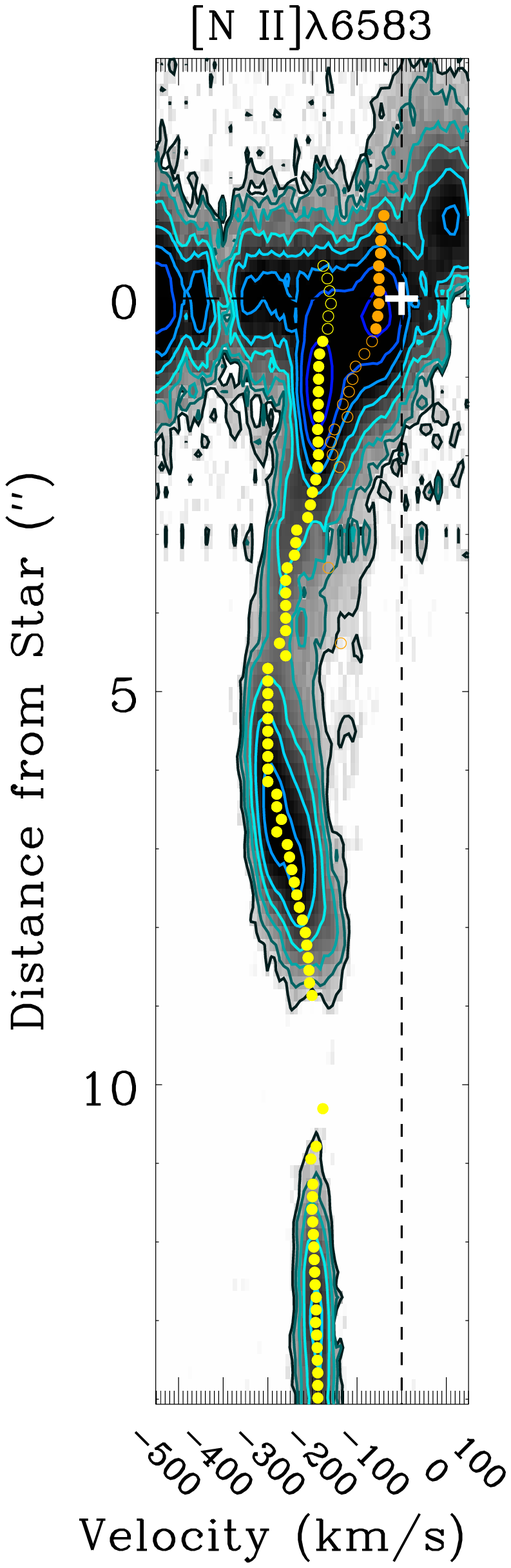} 
\hspace*{-0.2cm}
\includegraphics[width=0.2\textwidth]{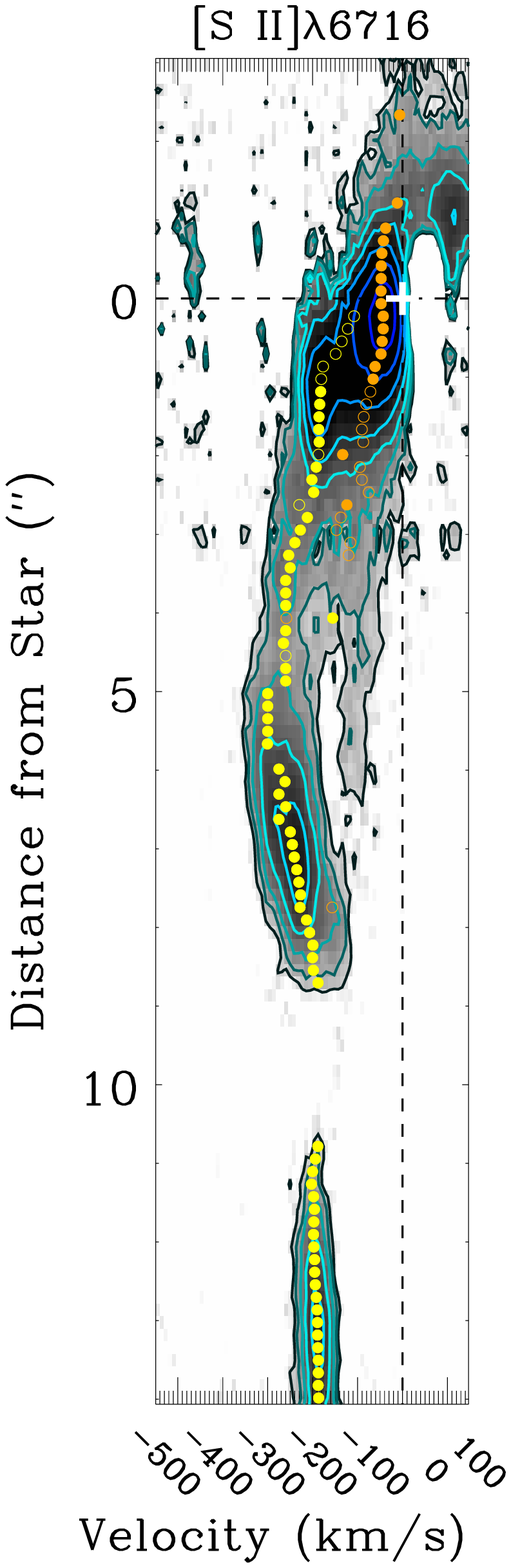} 
\hspace*{-0.2cm}
\includegraphics[width=0.2\textwidth]{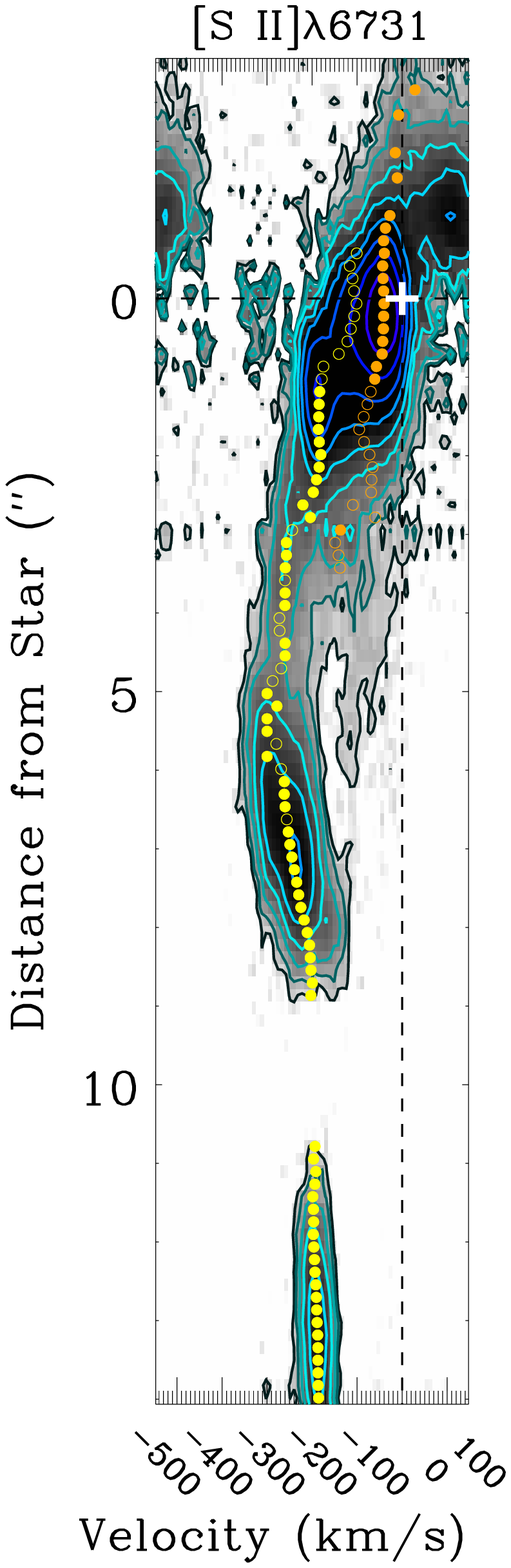} 
\hspace*{-0.2cm}
\caption{Selected PV diagrams of bright emission lines from 
         the VLT/X-Shooter VIS spectra, showing [O {\sc i}] $\lambda6300$, 
         H$\alpha$, [N {\sc ii}] $\lambda6583$, 
         [S {\sc ii}] $\lambda6716$, and [S {\sc ii}] $\lambda6731$. 
         The line profiles have been decomposed into high-velocity (yellow)
         and low-velocity (orange) components. At each position, the fitted 
         velocity centroids of the components are overlaid on the PV diagrams,
         with the stronger component in filled circles and the weaker component
         in open circles. In all the PV diagrams, the contours start from $3\sigma$ 
         ($1\sigma=3\times10^{-18}$ erg\,s$^{-1}$\,cm$^{-2}$\,\AA$^{-1}$)
         and increase by factors of 2.}
\label{PV_VIS}
\end{figure*}

The 2010 VLT/X-Shooter spectra provide a larger field of view along the blueshifted jet.
This data set provides an examination of ejection history after the 1999 HST/STIS observation.
The evolution of the microjet and the high-velocity bow-shaped knot can be seen 
in the VLT/X-Shooter spectra. In order to match the features between the two observations,
we assume a mean proper motion of $\sim 0\farcs28\,{\rm yr}^{-1}$ for the knots in the blueshifted jet
\citep{DGTau_RadioJet}. During the 11 years between the taking of the HST/STIS and the
VLT/X-Shooter spectra any cohesive features will have moved outward by $\sim 3\arcsec$.
In the following discussions we apply this offset in attempting to align features seen 
in the spectra and images.

     Figures \ref{PV_UVB} and \ref{PV_VIS} show the 
continuum-subtracted PV diagrams of bright optical lines of the 
jet up to $\sim 14\farcs2$ from the star. 
Three distinct emission regions can be identified from
the PV diagrams. (1) The strongest peak occurs within $\sim0\farcs5$ of the stellar 
position. The emission traces the flow with a trend of increasing speed and 
decreasing line intensity up to a local minimum at $\sim 4\arcsec$ from the star.
This coincides with the region interior to knot {\sl k3} \citep{DGTau_RadioJet} found 
in the [Fe {\sc ii}] wind \citep{Takami02,Pyo03} observed in 2001 to 2002, and may also 
correspond to the innermost $\lesssim 1\arcsec$ region of the 1999 STIS spectra 
\citep{Maurri14} and the [Fe {\sc ii}] microjet observed in 2005 \citep{AA11,White14}.
We identify the first emission peak and the extension as the ``star $+$ microjet'' 
region. (2) The second region peaks at $\sim 6\farcs5$ and may be identified with the
bow-shaped knot observed at $3\farcs6$ in 1998 \citep{LCD00}, denoted as knot B by
\cite{EM98}. The knot A \citep[or knot {\sl k2} in][]{DGTau_RadioJet} identified 
at $0\farcs93$ in 1998 \citep{LCD00} and $1\farcs3$ in 1999 \citep{Maurri14} appears
to have dissipated by 2009. It may have contributed to the velocity maximum
observed at $\sim 5\farcs5$. We denote this region as ``knot A$+$B.'' (3) The 
third region, peaked at $\sim 13\farcs2$, is truncated by the edge of the detector.
We identify this region as ``knot C.''

Overall, the kinematic features of emission lines of the different 
species are similar, with slight differences in the shapes and widths of the lines.
In the ``star $+$ microjet'' region, all species have broad 
line profiles that show either double emission peaks or a prominent
peak blended with a weaker ``shoulder.'' 
We decomposed the lines as in Section \ref{HSTobs}, 
and have overplotted the fitted velocity centroids
in Figures \ref{PV_UVB} and \ref{PV_VIS}, with the
HVC in yellow and the LVC in orange. For positions
that require double-Gaussian fitting, velocity centroids of the stronger
component are shown as filled circles and those of the weaker component 
are shown as open circles.

      The LVC is confined to within $\sim 3\arcsec$ of
the star and is less extended than the HVC. For all species, the LVC appears stronger than the HVC
within the innermost $1\arcsec$. The HVC becomes dominant outside $1\arcsec$
and extends beyond $3\arcsec$. Some of the high-ionization and high-density
lines, including [Ne {\sc iii}] and [O {\sc iii}], are undetected beyond
$\sim3\arcsec$ and reappear after $\sim5\arcsec$. 
The HVC velocity centroids gradually increase from $-180$ \kmps\ at $1\arcsec$ to 
$-230$ \kmps\ at $3\arcsec$. The velocity maxima
of $\sim -260$ \kmps\ are reached near the beginning of knot A at $\sim5\arcsec$ and 
decrease steadily to $\sim -170$ \kmps\ at the end of knot B at $\sim9\arcsec$.
At knot C, the jet has a steady velocity of $\sim -200$ \kmps.
Table \ref{tab_xsht} summarizes the kinematic properties of 
several forbidden emission lines obtained from Gaussian decomposition of
the 2010 VLT/X-Shooter spectra.

     The X-Shooter spectra show that most of the permitted line
profiles are centered at the position and systemic velocity of DG Tau.
Permitted lines that may be related to accretion flow, such as 
H {\sc i}, He {\sc i}, and Ca {\sc ii}, are broad, extending up to 
$\pm450$ \kmps. For some of the lines, such as H {\sc i} Balmer series 
(H$\alpha$ to H$\zeta$), Ca {\sc ii} H and K, and He~{\sc i} (specifically 
$\lambda5876$ and $\lambda7065$), profiles are asymmetric with an extension to the blue.
These lines also show spatially resolved emission from knot A$+$B,
where the kinematic structures are consistent with
forbidden lines that trace the jet emission.

\begin{table}[ht]
\begin{deluxetable*}{lccccccc}
 \tablewidth{0pt}
 \tablecaption{Kinematic Properties from the 2010 VLT/X-Shooter Spectra\label{tab_xsht}}
 \tablehead{
  \colhead{} & \colhead{[Ne {\sc iii}] $\lambda3869$} &
               \colhead{[S {\sc ii}] $\lambda4068$} & 
               \colhead{[O {\sc iii}] $\lambda5007$} & 
               \colhead{[O {\sc i}] $\lambda6300$} &
               \colhead{[N {\sc ii}] $\lambda6583$} &
               \colhead{[S {\sc ii}] $\lambda6731$} \\
  \colhead{Component (Region)} & \multicolumn{6}{c}{median $\pm$ std dev. (\kmps)}
 }
\startdata
LVC (microjet) & $-69.32\pm23.6$ & $-67.33\pm7.46$ & $-87.97\pm17.9$ & $-43.11\pm10.5$ & $-118.0\pm48.0$ & $-68.63\pm23.9$ \\
HVC (microjet) & $-187.5\pm11.8$ & $-187.7\pm23.6$ & $-189.6\pm14.8$ & $-182.7\pm29.2$ & $-187.0\pm14.0$ & $-184.3\pm27.6$ \\
HVC (A$+$B) & $-268.0\pm24.5$ & $-258.5\pm25.4$ & $-272.5\pm15.8$ & $-258.1\pm34.9$ & $-260.0\pm35.0$ & $-260.0\pm31.9$ \\
HVC (knot C) & $-187.5\pm13.2$ & $-188.2\pm7.64$ & $-186.0\pm6.60$ & $-192.1\pm5.13$ & $-193.0\pm10.1$ & $-190.5\pm4.76$ 
\enddata
\end{deluxetable*}
\end{table}

\subsection{Properties of the [Ne {\sc iii}] $\lambda3869$ Velocity Components} \label{NeIII}

\begin{figure*}
\epsscale{0.50}
\plotone{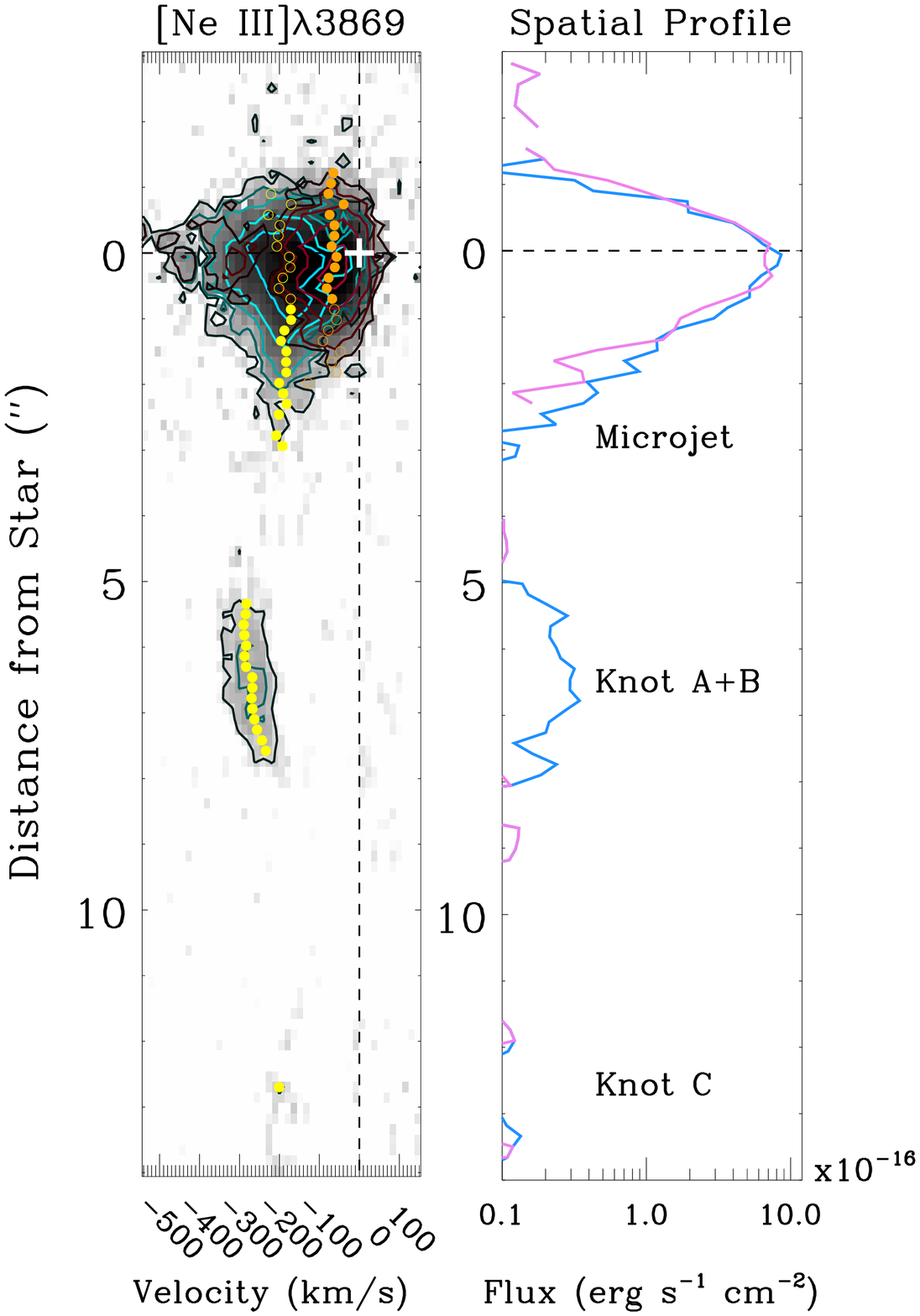}
\epsscale{0.45}
\plotone{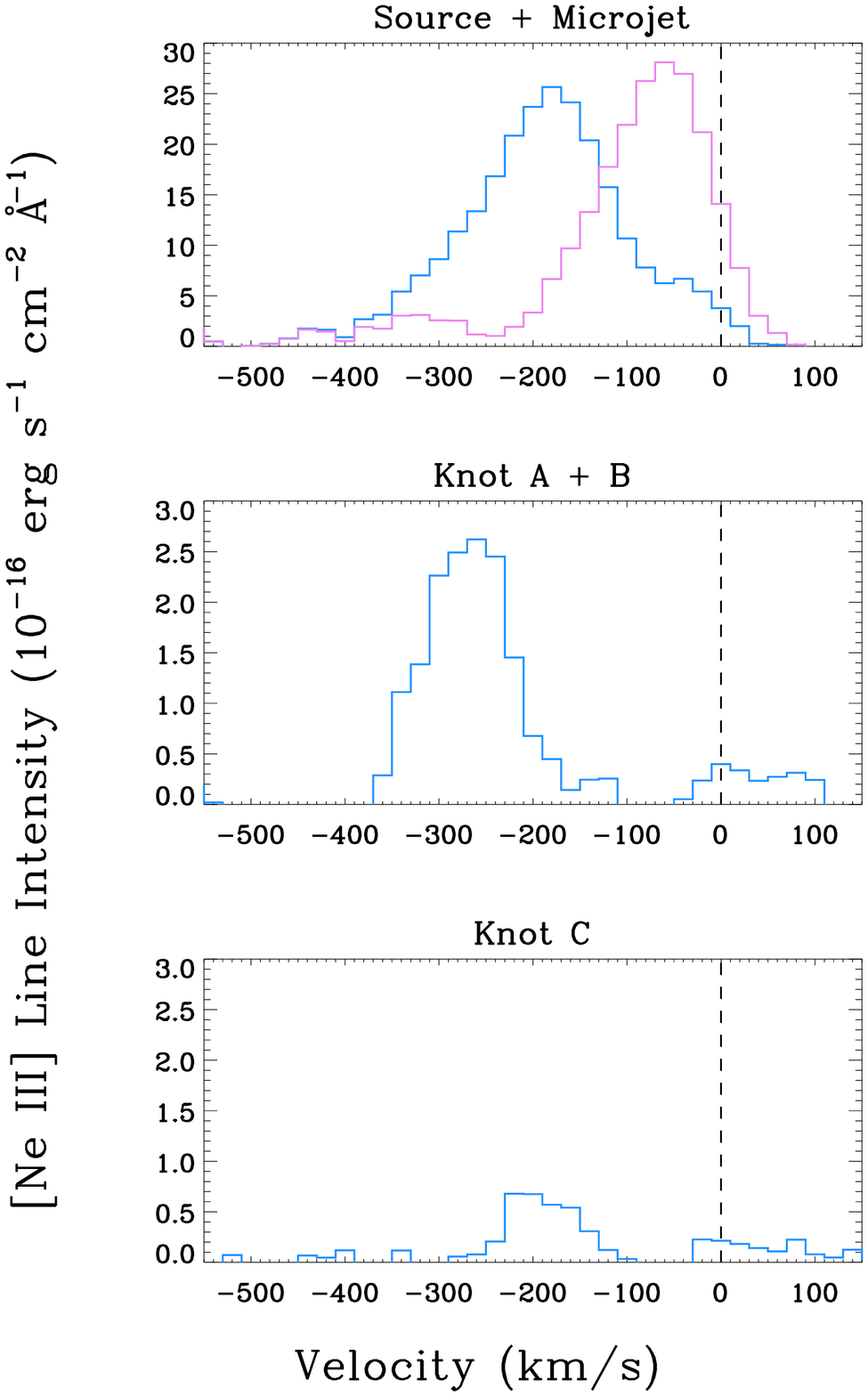}
\epsscale{1.00}
\caption{Kinematic properties of velocity-decomposed [Ne {\sc iii}] $\lambda3869$ 
         emission are shown in a PV diagram (left panel), a spatial profile along
         the jet axis (central panel), and line profiles at specific spatial
         positions (right three panels). In the PV diagram, the decomposed 
         HVC and LVC are shown by green and red contours, respectively. The 
         yellow symbols represent the HVC velocity centroids and orange symbols
         represent the LVC velocity centroids; the filled symbols indicate the 
         stronger component at each specific position whereas the open symbols
         indicate the weaker component, as in Figure \ref{PV_UVB}.
         The spatial profile is shown on a logarithmic scale to accommodate the 
         large contrast between the innermost 3\arcsec emission and the outer knots.}
\label{XS_NeIII_SpecSpatProf}
\end{figure*}

       Figure \ref{XS_NeIII_SpecSpatProf} shows the kinematic properties
of the velocity-decomposed [Ne {\sc iii}] $\lambda3869$ emission as PV
diagrams, spatial profiles, and line profiles at specific spatial positions
along the jet. [Ne {\sc iii}] $\lambda3869$ is detected toward the innermost 3\arcsec\
(star $+$ microjet) and between 5\arcsec\ and 8\arcsec\ (knots A and B),
and is marginally detected at knot C. The integrated 
fluxes for the three regions are $10.82\pm0.09\times10^{-15}$, $5.19\pm0.26\times10^{-16}$,
and $1.25\pm0.26\times10^{-16}$ erg\,s$^{-1}$\,cm$^{-2}$, respectively.

\begin{figure}
\plotone{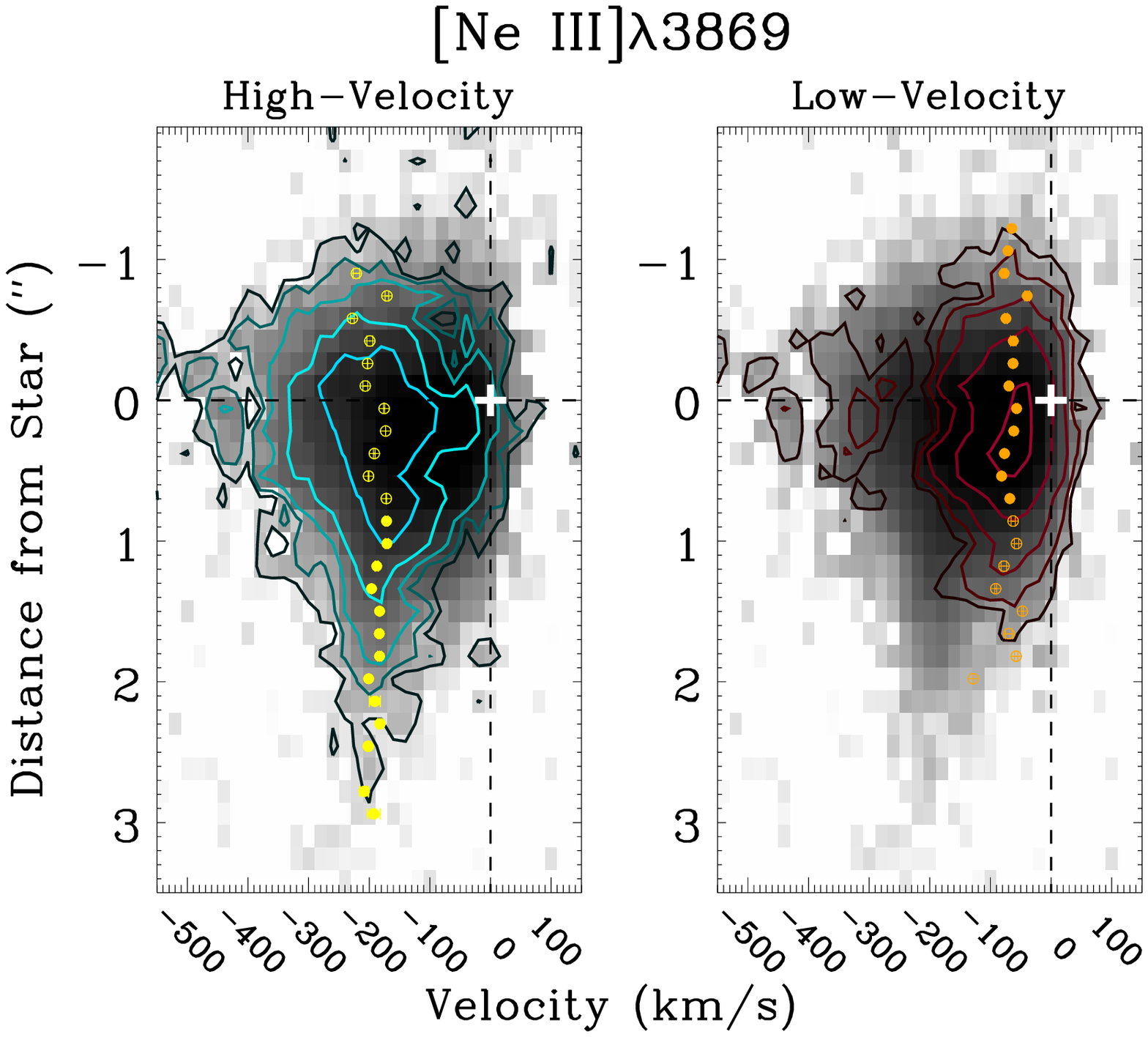}
\caption{Velocity-decomposed PV diagrams of {[Ne {\sc iii}]} $\lambda3869$
         from the innermost 3\arcsec\ blueshifted microjet. The left and 
         right panels show the HVC and LVC PV diagrams, respectively. Both
         contours start from $3\sigma$ ($\sigma = 3.0\times10^{-18}$ 
         erg\,s$^{-1}$\,cm$^{-2}$) and increase by factors
         of 2. The background grayscale maps are the original PV diagrams
         without velocity decompositions. The Gaussian-fitted velocity
         centroids are overlaid. The filled symbols indicate the stronger
         component at that spatial position and the open symbols indicate the
         weaker component.}
\label{NeIII_PVDecomposed}
\end{figure}

       The decomposed [Ne {\sc iii}] $\lambda3869$ PV diagrams
(Figure \ref{NeIII_PVDecomposed}) show the velocity
components in the innermost 3\arcsec\ of the flow.
After decomposition, the overall peak at $\sim -100$ \kmps, 
$\sim 0\farcs2$ from the star, breaks into two peaks of similar 
intensities at $\sim -180$ \kmps\ for the HVC and $\sim -70$ \kmps\ for
the LVC. The HVC peak has a line width of $\sim 200$ \kmps\ at 
0\farcs2, larger than the LVC line width of $\sim 120$ \kmps\ 
at the same position. The LVC has relatively stable line widths within
to 1\arcsec, whereas the HVC shows decreasing line widths from the 
maximum value of $\sim 200$ \kmps\ to $\sim 100$ \kmps\ at $\sim 2\arcsec$.
There is a non-Gaussian wing extending to $\sim -500$ \kmps.
The total fluxes in the HVC and LVC components within $\pm 3\arcsec$ are
$5.3\pm0.1\times10^{-15}$ and $4.8\pm0.1\times10^{-15}$ erg\,s$^{-1}$\,cm$^{-2}$, respectively. 
The kinematic properties of the decomposed [Ne {\sc iii}] $\lambda3869$
emission along the regions in the jet are summarized in Table \ref{tab_neiii}.

\begin{table}[bt]
\begin{deluxetable*}{lcccc}
 \tablewidth{0pt}
 \tablecaption{Properties of the [Ne {\sc iii}] $\lambda3869$ Jet of DG Tau\label{tab_neiii}}
 \tablehead{
  \colhead{} & \colhead{} & 
  \colhead{Velocity Centroid} & \colhead{Velocity Width} & \colhead{Flux} \\
  \colhead{} & \colhead{} & 
  \colhead{\kmps} & \colhead{\kmps} & \colhead{erg\,s$^{-1}$\,cm$^{-2}$}
 }
\startdata
Microjet & LVC & $-69.32\pm23.6$ & $107.4\pm28.6$ & $4.8\pm0.1 \times 10^{-15}$ \\
         & HVC & $-187.5\pm11.8$ & $123.0\pm38.0$ & $5.3\pm0.1 \times 10^{-15}$ \\
Knot A$+$B &   & $-268.0\pm24.5$ & $103.6\pm14.1$ & $5.2\pm0.3 \times 10^{-16}$ \\
Knot C (part) && $-187.5\pm13.2$ & $100.7\pm88.8$ & $1.3\pm0.3 \times 10^{-16}$  
\enddata
\end{deluxetable*}
\end{table}

\section{On the Possible Origins of [Ne {\sc iii}] Emission in the DG Tau Jet} \label{discussion_neon}
\subsection{[Ne {\sc iii}] Microjets from Low-Mass Young Stars}

   DG Tau is one of the few low-mass YSOs known
to show bright [Ne {\sc iii}] $\lambda3869$\AA\ emission
in the arcsecond-scale microjet. The [Ne {\sc iii}] emission
shares kinematic properties with other forbidden
emission lines tracing the jet (e.g., [O {\sc i}], [S {\sc ii}], and 
[N {\sc ii}]), notably the two velocity components at $-180$ and $-70$
\kmps. However, because of its higher ionization potential and higher critical density, 
the emission is restricted to the innermost 3\arcsec\ microjet and the 7\arcsec\ knot.

The presence of [Ne {\sc ii}] and [Ne {\sc iii}] emission from the jet may 
require that high-energy photons be present 
either at the launching site or during the propagation of the jet.
Doubly ionized neon arises from either
ionization of valence electrons, which requires a total energy of 62.5 eV,
or the photoionization of K-shell electrons, which requires a
photon energy $>$0.903 keV. In the low-mass circumstellar
environment, valence electrons may be ionized 
by extreme ultraviolet (EUV) photons or by collisions in shocks with
shock speeds higher than 100 \kmps\ \citep{HRH87,HG09}. K-shell photoionization
must occur close to the star \citep{GNI07}, a copious source of keV X-rays.

As described in Section \ref{Intro}, the X-ray source DG Tau is bright,
spatially extended, and consists of at least three distinct components.
The strongest component, located at the nominal stellar position,
is likely thermal emission from hot ($\sim$20--30 MK) coronal 
gas confined in magnetic loops \citep{Gudel08}.
The soft extended X-ray emission is concentrated $\sim 5\farcs5$ from the star. 
Its proper motion of 0\farcs28\,yr$^{-1}$ suggests association with 
non-standing shocks in the jet \citep{Gudel12,DGTau_RadioJet}.
The third component, the soft source 0\farcs2 from the star,
exhibited no significant proper motion over a 2 yr period from 2004 to 2006 \citep{SS08}. 
The emission could be produced by $\sim 450$ \kmps\ shocks 
in the innermost dense jet \citep{GML09}, colliding stellar and 
magnetocentrifugal winds \citep[from the inner disk;][]{GLS14},
or reconnections of magnetic fields threading the jet \citep{Sch13}.

The first known [Ne {\sc iii}] microjets from low-mass YSOs were discovered 
toward Sz 102 \citep{Liu14}. The nearly edge-on orientation of the Sz 102 system presents
a different geometry for comparison with the jet of DG~Tau.
In Sz~102 the [Ne {\sc iii}] $\lambda3869$ emission appears unresolved within 
$\sim200$ au in the ground-based VLT/{\sc Uves} spectra and shows a broad
line profile with ``excess'' emission across the systemic velocity. This suggests
that the jet originates from a wide-angle wind close to the star. The neon 
may be ionized by large hard X-ray flares generated by reconnection events 
from field lines twisted by star--disk interactions within the regions 
encompassed by the wide-angle wind and may then be carried out through the wind flow. 
Further monitoring of the system would
be required to test the postulation since virtually nothing is known about hard X-ray 
flares or temporal variability of the [Ne {\sc iii}] line in Sz~102.

For DG Tau, we discuss the origins of its [Ne {\sc iii}] jet 
in the following subsections. In summary, we suggest that the neon ionization in the
innermost microjet may be largely attributed to the photoionization by a 
series of flares from a bright hard X-ray source and the slow recombination
of the innermost hot jet heated through either jet shocks, recollimated stellar
wind shocks, or magnetic reconnection. A contribution from shock ionization 
in the jet is still possible, though the fraction of shock dissipation must be 
small in order to be consistent with the inferred soft X-ray luminosity. The 
detection of the outer [Ne {\sc iii}] knot would favor strong shock ionization in the jet,
which may also account for the soft X-ray extension along the jet axis.

\subsection{Ionization of the Inner 3\arcsec\ Microjet: Shocks and Soft X-rays?}

     Shocks can be an efficient ionization source for neon
when the postshock temperature reaches the order of megakelvin, above
which collisional ionization is important. 
At several megakelvin, thermal gas emits mainly in soft X-rays which peak at $< 1$ keV. 
To reproduce the spectral energy distribution of the 0\farcs2 soft X-ray component at 
$\sim 2$ MK requires a shock speed $v_{\rm s} \approx 480$ \kmps, 
gas temperature of $\sim 0.3$ keV, and volume emission measure of $\sim 10^{52}$ cm$^{-3}$ 
\citep{GML09}. To satisfy these constraints, the cooling length and cross section 
of the modeled shock must both be of the order of au, 
with the mass flux through the shock of the order of 
$10^{-11}$ -- $10^{-10}$ $M_\odot$\,yr$^{-1}$ \citep{GML09}. The ratio between the 
mass flux in the shock and the total mass-loss rate of DG Tau ($\sim 10^{-7}$ 
$M_\odot$\,yr$^{-1}$), $\sim 10^{-4}$ -- $10^{-3}$, may be regarded as the 
fraction of shock dissipation $f_{\rm sh}$ as in \cite{HG09}. 
One can compare the theoretical and observed [Ne {\sc ii}] 12.81\micron\ 
flux using Equation (33) of \cite{HG09}.
Putting $n_0 \approx 10^5$ cm$^{-3}$, $v_{\rm s} \approx 480$ \kmps, and $f_{\rm sh} 
\approx 10^{-3}$, the predicted luminosity is $10^{-6} L_\odot$, two orders of magnitude 
fainter than the observed {\it Spitzer} value \citep[$\sim 2\times10^{-4} L_\odot$,][]{Gudel10}. 
Although the {\it Spitzer} observation is spatially unresolved, the expected 
contribution from the disk atmosphere would be at the level of $\sim10\%$ \citep{Gudel10}.
The bulk of the [Ne {\sc ii}] 12.81\micron\ flux must arise via a different mechanism.

Pure shock models, whether planar shocks or bow shocks, make predictions of line ratios
\citep[e.g.,][]{HRH87} that can be compared with our data. 
The main adjustable parameters are the shock velocities and preshock densities.
High-ionization forbidden lines, such as [Ne {\sc iii}] 
$\lambda3869$ and [O {\sc iii}] $\lambda5007$, and UV permitted lines, such as
C {\sc iv} $\lambda1550$, appear only when shock velocities exceed $\sim 100$ \kmps. 
For forbidden emission lines, the relative intensities increase as shock velocity 
increases, until collisional de-excitation becomes important at velocities higher 
than $\sim300$ \kmps. For example,  as shock velocities increase from 100 to 300 \kmps,
[O {\sc iii}]/H$\beta$ decreases from $\sim4.0$ to $\sim2.5$, and
[Ne {\sc ii}] 12.81\micron/H$\beta$ decreases from $\sim3.0$ to $\sim1.5$.
The relative flux of UV permitted lines decreases
as the Balmer intensity increases since more ionizing photons in 
the shock front are absorbed by H atoms: C {\sc iv}/H$\beta$, 
for example, decreases from $\sim20.0$ to $\sim5.5$.
The [Ne {\sc iii}]/H$\beta$ ratio, on the other hand, increases by 
$\sim20$\% as shock velocity increases from 300 \kmps, and approaches $\sim 1.0$. 
This results in decreasing line ratios of C {\sc iv}/[Ne {\sc iii}], 
[O {\sc iii}]/[Ne {\sc iii}], and [Ne {\sc ii}]/[Ne {\sc iii}] for shock speeds exceeding 300 \kmps.
The C {\sc iv}/[Ne {\sc iii}] ratio approaches $\sim 10$, the [O {\sc iii}]/[Ne {\sc iii}]
ratio approaches 3, and [Ne {\sc ii}]/[Ne {\sc iii}] approaches 1 for high-velocity shocks. 
The observed ratios for the inner 3\arcsec\ jet in the HVC are 
\citep[dereddened with $A_V \approx 0.55$ as in][]{Sch13L}
[Ne {\sc iii}]/H$\beta$ $\sim 0.15$, [O {\sc iii}]/H$\beta$ $\sim 0.12$, and 
C {\sc iv}/H$\beta$ $\sim 1$, which give C {\sc iv}/[Ne {\sc iii}] $\sim 8$,
[O {\sc iii}]/[Ne {\sc iii}] $<1$, and [Ne {\sc ii}]/[Ne {\sc iii}] $\sim 24$.
The C {\sc iv}/[Ne {\sc iii}] ratio appears to be similar to the ratio in the
shocked gas but the [O {\sc iii}]/[Ne {\sc iii}] and [Ne {\sc ii}]/[Ne {\sc iii}]
ratios are not consistently reproduced.
The high [Ne {\sc ii}]/[Ne {\sc iii}] value, as produced by
high [Ne {\sc ii}]/H$\beta$ and low [Ne {\sc iii}]/H$\beta$, occurs more readily
in the mild shock condition when the shock speed is lower than $\sim 100$ \kmps. However,
this would be inconsistent with the assumption that the soft X-rays are produced by
strong shocks in the jet. Discrepancies from the crude comparison of 
line ratios can be due to the different spatial coverages of the line species that 
cannot to be resolved from spatially summed spectra. Spatially resolved
spectra and a shock model detailing the geometry are needed to account for possible spatial 
variations of the flux ratios.

An alternative explanation for the 0\farcs2 stationary X-ray source is the 
collimation shock where the tenuous stellar wind collides with the inner boundaries of 
the magnetocentrifugal winds emerging from
the inner region of the circumstellar disk \citep{GLS14}. A soft X-ray
source produced by this stationary shock lies outside the jet and can irradiate it.
The X-ray ionization rate of doubly ionized neon
depends on the X-ray luminosity and the branching ratio of the production rate from
neutral to doubly ionized neon.
These soft X-rays are an order of magnitude fainter than the harder coronal X-rays \citep{Gudel08}.
The branching ratio required for the soft X-ray to doubly ionize neon is $\sim0.1$,
much smaller than $\sim 0.94$, that for the hard X-ray \citep{GNI07}.
Therefore, even if the soft X-ray irradiation contributes to
neon ionization, in this scenario the expected ionization rate will be more than
an order magnitude smaller than that provided through harder coronal X-ray irradiation.

\subsection{Ionization of the Inner 3\arcsec\ Microjet: Hard X-Rays and Flares?}

   The keV photons from the harder coronal X-ray source coincident with the star 
\citep{Gudel08,SS08} provide a more efficient source for ionizing the neon.
The large ionization flux from hard X-rays can produce a large fraction
of electrons. The large wind mass-loss rate of DG Tau 
($\dot{M}_{\rm w} \sim 10^{-7}$ $M_\odot$\,yr$^{-1}$) produces relatively dense regions in the jet.
Combining the two effects, sufficient collisional excitation can be maintained to explain 
the observed [Ne {\sc iii}] $\lambda3869$ flux.
By reproducing the observed mid-infrared
[Ne {\sc ii}] and [Ne {\sc iii}] luminosities of jet-driving low-mass YSOs, 
\cite{SGLL} demonstrated that both X-ray ionization and high mass-loss rate are
crucial to sustain a high collisional excitation rate.
In contrast to the strong shock scenario, photoionization by keV photons 
is the main ionization source \citep{GNI07}. Only mild shocks with speeds $\lesssim 50$ 
\kmps\ are needed in the jet to maintain the temperature and ionization fraction
for the formation of optical forbidden emission lines.
The observed [Ne {\sc iii}] $\lambda3869$ HVC flux, dereddened using $A_V = 0.55$,
corresponds to a luminosity of $\sim 2.7\times10^{28}$ erg\,s$^{-1}$.
The model of \cite{SGLL} would require an averaged X-ray luminosity
$L_{\rm X} \lesssim 10^{31}$ erg\,s$^{-1}$, or about 
an order of magnitude larger than observed, to explain the [Ne {\sc iii}] luminosity, 
for the inferred $\dot{M}_{\rm w}$.   
The apparent discrepancy can be accounted for by the fact that the X-ray 
luminosity adopted in the model refers to the intrinsic, averaged value of
X-ray luminosity. The intrinsic X-rays for T Tauri stars with mass accretion rate higher than
$10^{-8}\,M_\odot\,{\rm yr}^{-1}$ are inferred to be deficient by more than an order of magnitude
\citep{Telleschi07,Bustamante16}. The averaged value may also be elevated when flares are 
taken into account.

Stellar flares could produce the required ionizing flux for 
the observed neon ionization and [Ne {\sc iii}] $\lambda3869$ intensity.
X-ray surveys of low-mass young stars in star clusters and associations 
show that impulsive flares typically occur 
on a timescale of weeks and rise to average luminosities 
of $\sim 6\times10^{30}$ erg\,s$^{-1}$ in a day or two \citep{Wolk05}. 
Larger but rarer flares of $\sim 100$ MK with luminosities up to 
$\sim 10^{32}$ erg\,s$^{-1}$ have been observed
\citep{Imanishi01,Grosso04,Favata05,Getman08a,Getman08b}. 
These strong flares may be associated with reconnections in closed loops of 
stellar magnetic fields extending to several stellar radii \citep{Favata05}.
Interactions between the star and its circumstellar disk
can also induce reconnections and flares \citep{SHGL}. 

The flare frequency of DG Tau is poorly known. DG Tau has been observed by
{\it Chandra} in several short exposures (each $\sim 30$ ks) in 2004--2006, and 
three long exposures (each $\sim 120$ ks) in 2010 January \citep{Gudel12}. The long
exposures captured two flare events within three days with durations of 30 and 50 ks,
respectively. The strongest flare on 2010 January 5, just two weeks before our 
VLT/X-Shooter observation, shows an increase of five times the average luminosity
of the X-ray emission at energies above 1 keV. Such a large X-ray luminosity,
$\sim 5\times10^{30}$ erg\,s$^{-1}$, may contribute a significant fraction of the neon
ionization. At densities below the critical density of [Ne {\sc iii}], frequent X-ray flares
could maintain the observed emission measure. Since we do not have a good record
of the [Ne {\sc iii}] flux, further extrapolations and predictions are dangerous.

If neon is ionized primarily by hard X-ray flares,
the [Ne {\sc iii}] emission will peak close to the ionization source, 
where the ionization rate and collisional excitation rate are the maximum. The line 
intensity will gradually decrease as the flow propagates due to recombination
of the doubly ionized neon. The observed [Ne {\sc iii}] $\lambda3869$ spatial profile along 
the jet axis is basically consistent with this scenario of ionization and 
recombination (Figure \ref{XS_NeIII_SpecSpatProf}).

In this scenario, how far the doubly ionized neon can extend along the jet depends on how
long the ionization remains frozen-in along the flow if no strong secondary 
ionization source occurs during propagation of the jet. 
The observed spatial extent of the emission may be compared
against the expected extent that is determined by the estimated values of
the recombination timescale and the flow timescale. The 
high-velocity [Ne {\sc iii}] microjet originates close to the source and 
extends up to 2\arcsec--3\arcsec. This is among the shortest extent of any of the 
observed forbidden HVC emission from the DG Tau microjet. 
The radial velocity of the HVC is $\sim -180$ \kmps. If a line-of-sight 
inclination of $\sim 40\degr$ is assumed \citep{EM98,Sch13L}, the derived 
proper motion on the plane of sky is 0\farcs23\,yr$^{-1}$ at the distance of 
140 pc. If no further ionization occurs in the flow, the neon gas would require 
a recombination timescale of
$\sim 10$ yr in order to explain the extended emission. The recombination timescale of 
[Ne {\sc iii}], $t_{\rm rec}$ , scales as $(n_e\alpha(T))^{-1}$. Using the coefficient 
$\alpha(T) \approx 1.7\times10^{-12}(T/10^4\,{\rm K})^{-2/3}$ \citep{GNI07,SGLL}, the 
recombination timescale is of the order of $\sim 1$ yr if $T \lesssim 3\times10^4$ K
and $n_e \approx 3\times10^4$ cm$^{-3}$ as implied by the red [S {\sc ii}] doublet
and [O {\sc i}]/[N {\sc ii}] \citep[e.g.,][]{Cof08,Maurri14}. Other sources
to maintain the ionization or temperature may be needed to increase the 
recombination timescale in order to better interpret the observation.

   The soft X-ray source at $\lesssim 0\farcs2$ from the star
\citep{Gudel08,SS08} may alternatively help to maintain the frozen-in 
ionization of the flow by contributing to jet heating.
The hot ($T \gtrsim 10^5$ K) C\,{\sc iv} jet, detected in 2011
somewhat downstream from the soft X-ray source, was considered to be a
consequence of the heating \citep{Sch13L,Sch13}. The peak emission is at 
$\sim -180$ \kmps, consistent with the velocity of HVC [Ne\,{\sc iii}] emission obtained 
$\sim 1$ yr earlier. The association of the C\,{\sc iv} jet with 
the [Ne {\sc iii}] HVC suggests that the lower-temperature emission of the 
jet may be a continuation of the hot jet. This hot portion of the jet can help 
drag out the recombination timescale to $\sim 10$ yr
and may account for the extended emission of the [Ne\,{\sc iii}] line.
In this scenario, the jet is propagating through the 
apparently stationary local heating region.
The heating can be due either to a jet shock of speed $\sim 450$ \kmps\ \citep{GML09} 
or to a recollimation shock of a stellar wind with velocity of $\sim 840$ \kmps\ obliquely
hitting a magnetocentrifugal wind \citep{GLS14}. The heating may also produce soft 
X-ray photons that partially contribute to C\,{\sc iv} and [Ne\,{\sc iii}] emission. 
Heating may also come from magnetic recombination of the field lines threading 
the jet \citep{Sch13L}. Specifically, the magnetic recombination in the jet fields 
may produce keV photons and provide additional ionization of neon gas in the jet.

For the ionization of the inner microjet, both the mechanisms of 
strong shocks and hard X-rays have their respective applicabilities and challenges.
Shock ionization can account for the neon ionization and the soft X-ray emission
0\farcs2 from the star if the shock speed reaches $\sim 480$ \kmps. However, the 
fraction of shock dissipation that matches the X-ray luminosity may be two orders
of magnitude lower than that needed to sustain the observed flux of [Ne {\sc ii}]
and [Ne {\sc iii}] and does not match the [Ne {\sc ii}]/[Ne {\sc iii}] ratio predicted
in shock models. The hard coronal X-rays from the close vicinity of the star can
overcome the threshold energy for neon photoionization, but an elevated average hard
X-ray luminosity greater than the observed value would be required to penetrate the outflowing
gas for the inferred mass-loss rate. The observed flares from DG Tau can provide
the elevated luminosity to fulfill this requirement. We further postulate the source
of the flares that can match the requirement in Section \ref{discussion_vc}.

\subsection{Ionization of the 6\farcs5 Knot: Shocks in the Jet?}

   The [Ne {\sc iii}] HVC is absent between 3\arcsec\ and 
5\farcs5, which is where knot A$+$B appears. The spatial discontinuity suggests that the 
physical conditions in the jet become less favorable for formation of the neon line 
such that the flow timescale cannot compete with the recombination timescale at 
distances larger than $\sim 3\arcsec$. At the outer knot, either the neon gas is 
reionized to a doubly ionized state or the electron density increases such
that the excitation conditions favor [Ne {\sc iii}] emission, or both.
From the PV diagram, the [Ne {\sc iii}] emission
decelerates from $\sim -290$ \kmps\ at 5\farcs5 to $\sim -210$ \kmps\ at
8\arcsec. When compared with the arcsecond-resolution {\it HST}/STIS long-slit spectra, 
this region corresponds to the knots B0 and B1 that also show a clear velocity jump 
from $\sim -290$ \kmps\ to $\sim -210$ \kmps\ \citep{Maurri14}. Taking into 
account for the inclination angle, the velocity jump is $\sim 100$ \kmps, 
which may be sufficient for in situ neon ionization if the velocity jump is 
interpreted as a strong shock in the jet flow \citep{HG09}. 

   The shocks responsible for the neon ionization in the outer knot may also 
result in the extended X-ray emission of DG Tau.
Adjacent to the [Ne {\sc iii}] knot at $\sim 6\farcs5$, the X-ray knot is at 
$\sim 5\farcs5$ in the 2010 {\it Chandra} image.
It was discovered at 4\farcs3 in 2005, and thus shows a proper motion 
of $\sim 0\farcs27$\,yr$^{-1}$ similar to those of other optical and infrared knots 
\citep{Gudel08,Gudel12,DGTau_RadioJet}. Tracing this proper motion back in time,
the X-ray knot would have appeared in the $\sim 2\farcs4$ interknot region between knots 
A1 and B0 in the 1999 {\it HST}/STIS spectra \citep{Maurri14} and the 1998 CFHT/OASIS
spectra \citep{LCD00}. Comparisons of the PV diagrams between the 1999 STIS 
spectra and the 2010 X-Shooter spectra show that the maxima of the velocity centroids 
have decreased from $\sim -360$ \kmps\ to $\sim -290$ \kmps, which also coincides with
the velocity jump between A1 and B0 \citep{Maurri14}. 
The interknot region may produce adequate ionization from shocks 
but the higher densities downstream are more favorable for line production.
The velocity difference of $\sim 100$ \kmps\ would be deprojected to $\sim 130$
\kmps\ for an inclination $i\approx 38\degr$.
The observed velocity jump appears to be less than the required
$\sim 380$ \kmps\ \citep{Shock_ARAA93} if shock is the main heating source for the
2.7 MK X-ray knot. A magnetic field carried by the flow may be working
to reduce the shock strength and provide additional heating \citep{Gudel08}, although the
field structures in the flow need further observations and modeling to be confirmed.

\section{Temporal Variations in the HVC of the DG Tau Jet} \label{discussion_vc}
\subsection{The Double Velocity Components of DG Tau}

DG Tau is known to have distinct double-peaked velocity profiles in its jet emission \citep{HEG95}.
\cite{SB93} used long-slit spectra to show the different properties of the two
components: a compact LVC at $-50$ \kmps\ appears close to the source, and 
a HVC at $-260$ \kmps\ is offset from the source by $\sim 0\farcs2$ 
and extends beyond 2\arcsec\ with a lower velocity of $\sim -200$ \kmps. 
Among the line species detected in the spectra,
[O {\sc i}] $\lambda6300$ appears in both components, [N {\sc ii}] $\lambda6583$
is dominant in the HVC, and [S {\sc ii}] $\lambda6731$ is dominant in the LVC. 
From the 1999 {\it HST}/STIS spectra, we again show that both velocity components
exist in the first 0\farcs7 of the microjet and that different
line species dominate in different velocity components.
In the near infrared, DG Tau also possesses [Fe {\sc ii}] 1.644\micron\ emission that can
be decomposed into a more compact and transversely broader LVC, and a more extended and
transversely narrow HVC \citep{Pyo03,AA11,White14}. The [Fe {\sc ii}] HVC is typically 
$\sim-200$ \kmps, and the LVC is $\sim-100$ \kmps. The velocity centroids of the [Fe {\sc ii}] 
LVC are similar to those of [N {\sc ii}] and faster than those of [O {\sc i}] and [S {\sc ii}].
Yet another molecular wind component, detected through H$\rm_2$, appears as a wide-angle,
low-velocity ($\sim -15$ \kmps) emission close to the source \citep{AA14,White14}.
The two-component profiles in the atomic lines, observed through spatially resolved 
spectra taken from 1980 to 2010, appear to have persisted for several decades in the DG~Tau jet.

    That the different line species are dominant in different velocity components is often 
considered to be consistent with an ``onion-shaped'' outflow structure \citep{Bac00,Pyo03}. 
In this scenario, the density and velocity structures, as well as excitation conditions, 
are layered across the jet axis. The species with higher excitation criteria (e.g., [N {\sc ii}]) 
appear in the central region in which the velocity and density are higher, whereas 
those with lower excitation conditions and lower critical densities (e.g., [S {\sc ii}]) 
appear in the outer region with lower velocities. 
On the other hand, this can also be achieved by a cylindrically stratified wide-angle wind,
similar to a steady-state X-wind. As described in \cite{Shang06,SGLL}, the divergent streamlines
contribute to the large line widths of the HVC, and axially concentrated density provides the 
bulk line emission and narrow appearance of the HVC, whereas the ambient material may be mildly
shocked to produce the LVC and the even slower molecular layer. For young flat-spectrum stars, 
the remaining envelope can provide the remnant ambient material. In the cases of DG Tau \citep{Testi02}
and HL Tau \citep{Cabrit96}, a remnant envelope perpendicular to the outflow has been detected, 
and low-velocity wide-angle H$\rm_2$ emission encompassing the star and the [Fe {\sc ii}] jet
is found \citep{Takami07,AA14,White14}.
Regardless of the interpretations, the emission of the HVC and LVC should be decomposed and 
separated because of their differences in the line forming regions. 

Through the analysis of two-Gaussian spectral decomposition, we show that the HVCs of 
[O {\sc i}], [S {\sc ii}], and [N {\sc ii}] emission lines can have systematic differences
in their respective velocity centroids. From the inner 0\farcs7 microjet in the 1999 STIS spectra,
[N {\sc ii}] shows a faster bulk radial velocity than [O {\sc i}] and [S {\sc ii}].
From the inner 3\arcsec\ microjet in the 2010 X-Shooter spectra, different species
have similar velocities. Explaining the different HVC centroids among species can be 
challenging for currently available models based on magnetocentrifugal wind. A steady-state
X-wind can successfully describe a predominantly single HVC that is uniform 
among species, such as in RW Aur A \citep{LS12} and in the 2010 X-Shooter spectra
of DG Tau, but would require more complex ionization and heating profiles for the jet 
in order to explain the differences shown in the 1999 STIS spectra of DG Tau.
The disk-wind model, with a layered velocity structure, also cannot fully
address this issue
\citep[see, e.g., the similarity of the synthetic {[O {\sc i}]} and {[S {\sc ii}]} spectra in][]{Rub14}. 
In either interpretation, the [N {\sc ii}] line emission traces the innermost dense region
close to the jet axis and the velocity centroid represents the flow emanating from the
innermost part of the disk. We regard velocity centroids traced by the [N {\sc ii}] emission 
as the representative speed of the DG Tau jet in the subsequent discussions on 
temporal variations of the velocity components.

\subsection{Temporal Variations of the Double Velocity Components}

\begin{figure*}
\plottwo{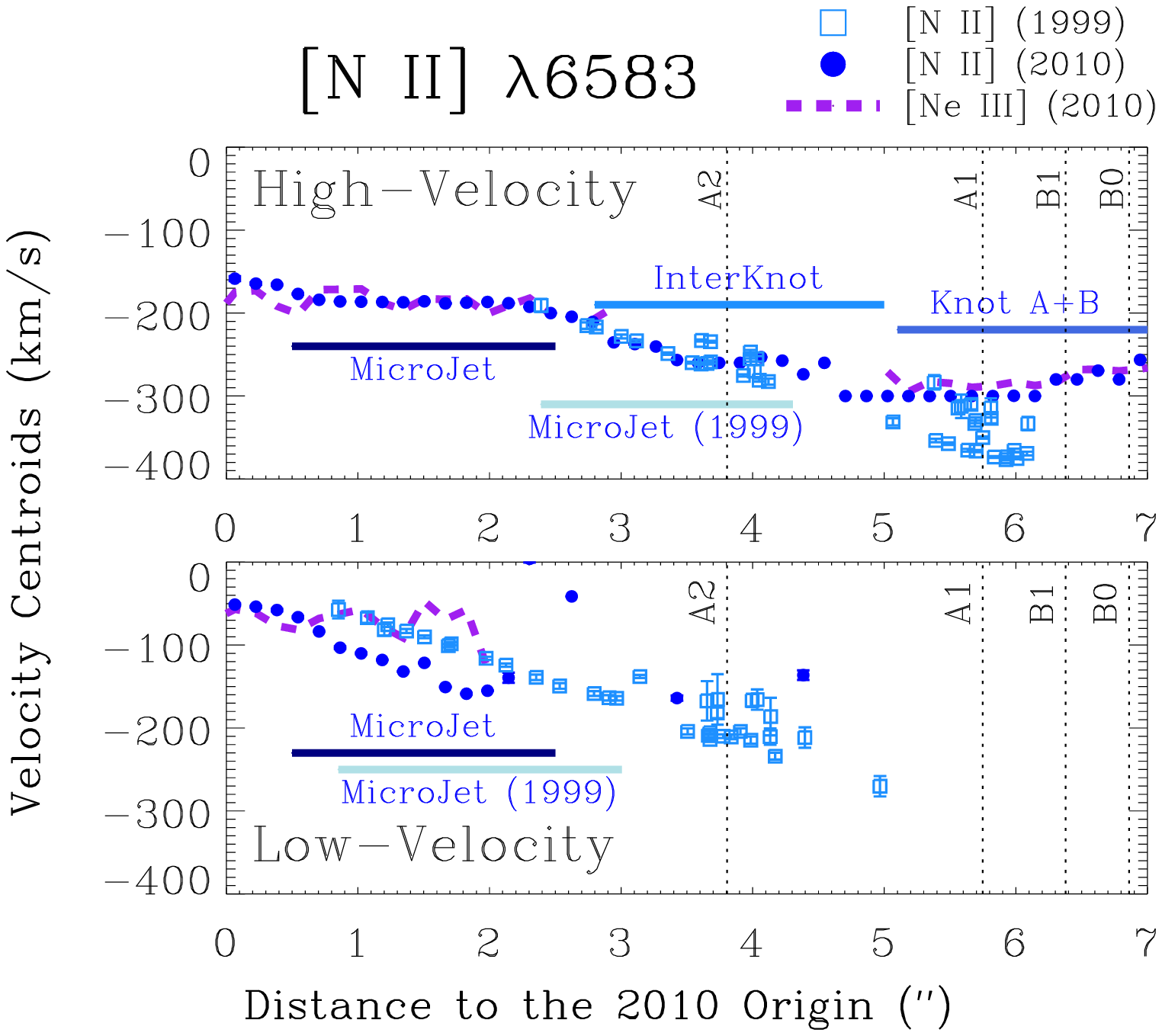}{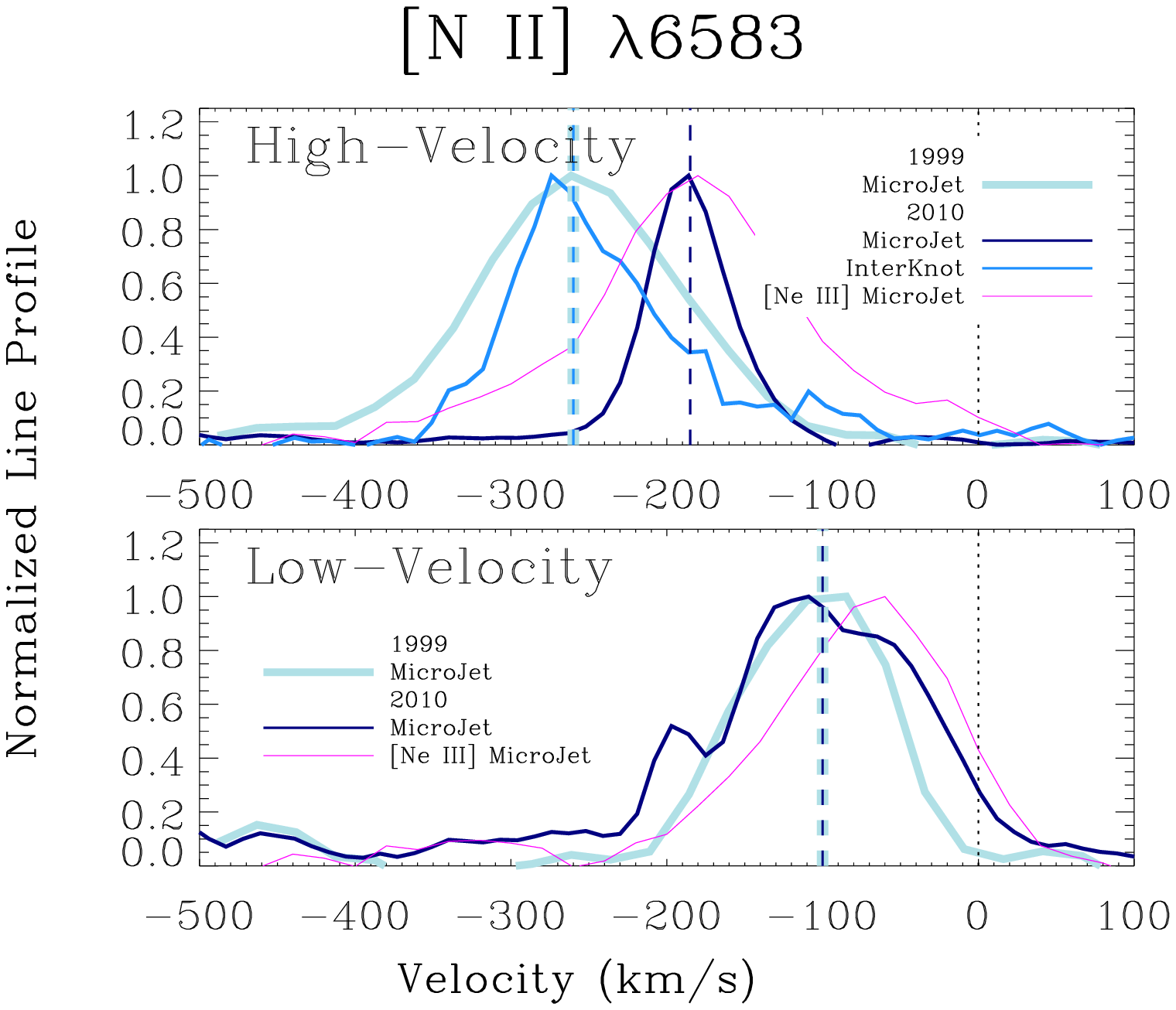} 
\caption{Variations of the [N {\sc ii}] $\lambda6583$ HVC and LVC centroids
         between the 1999 {\it HST}/STIS and 2010 VLT/X-Shooter spectra. 
         The left panels compare spatial variations of the velocity centroids 
         along the jet axis at the two epochs. The origin is set at the 
         2010 observational epoch, and the 1999 data are shifted to positions according to 
         their respective velocity centroids, adopting the 
         inclination angle \citep[$\sim 38\degr$,][]{EM98} and distance \citep[140 pc,][]{TLMR07}. 
         The vertical dashed lines indicate the positions of features defined in 
         \cite{Maurri14}, shifted according to their respective velocity centroids 
         obtained from the 1999 spectra. The 1999 (open squares) and 2010 (filled circles) 
         [N {\sc ii}] data are compared against [Ne {\sc iii}] $\lambda3869$ (thick dashed lines)
         from the 2010 data. The regions of ``MicroJet,'' ``InterKnot,'' and ``Knot A$+$B''
         are defined based on the [Ne {\sc iii}] features, and the ``MicroJet (1999)'' region is
         defined by the positions of the shifted 1999 velocity centroids. The right panels 
         compare the line profiles at the two epochs, normalized to their individual 
         maxima. The light thick lines are the velocity-decomposed profiles of 
         [N {\sc ii}] from the 1999 data, obtained by summing the spectra over the 
         ``MicroJet (1999)'' region. The strong thin lines are those 
         from the 2010 data; the darker profiles sum over the ``MicroJet'' region and the
         lighter profiles sum over the ``InterKnot'' region. [Ne {\sc iii}] $\lambda3869$ 
         profiles summed over the ``MicroJet'' region are shown for reference by thin purple lines.}
\label{optical_timevar_VC}
\end{figure*}

While they are persistent, both velocity components of the DG Tau jet have varied.
The most obvious differences can be seen 
by comparing the 1999 {\it HST}/STIS spectra and the 2010 VLT/X-Shooter spectra. 
As described above, the HVC centroids identified in the 1999 STIS spectra differ
among species (cf. Figure \ref{STIS_avgPV}), ranging from $\sim -150$ to $\sim -260$ \kmps.
Such differences seem to occur only in the innermost 0\farcs7 region in the 1999 STIS spectra.
On the other hand, previous ejecta that correspond to the region
of knots B0 and B1 show more uniform velocity centroids among [O {\sc i}], [S {\sc ii}],
and [N {\sc ii}] \citep{LCD00,Maurri14}. Post-1999 ejecta, e.g., those
corresponding to the inner 3\arcsec\ region in the 2010 X-Shooter spectra, also show
uniform velocity centroids among species, with a median value of $\sim -180$ \kmps. 
Understanding how the star transitions between these 
two phases and whether the phase with non-uniform velocity centroids will recur may 
require spectroscopic monitoring of the DG Tau jet.
Variations in the LVC are less obvious. Within the 
innermost 1\arcsec, the velocities at the two epochs appear consistent.

The comparisons between the 1999 {\it HST}/STIS spectra and the 2010 VLT/X-Shooter spectra
can be compared by matching their spatial and spectral resolutions.
The spectral resolution for both observations is $\sim 65$ \kmps. The main differences are 
the spatial resolution and the settings of the slit observations. The point spread function (PSF) 
of the {\it HST} optical instruments is defraction-limited to $\sim 0\farcs1$, whereas that of 
the VLT is seeing-limited to $\sim 1\arcsec$ at the time of the observation. The slit widths
match the PSF at observations, and the widths of 0\farcs1 and 1\farcs2 were used for STIS and 
X-Shooter, respectively. Although spatial resolutions cannot be matched, spatial coverages of
the two observations can be kept similar for the analysis. Therefore, the STIS spectra 
transversely averaged across 0\farcs52 were used when comparisons of velocity centroids 
were made and longitudally averaged line profiles from both spectra were used to compare the 
overall representative spectral features between the two observational epochs.

Figure \ref{optical_timevar_VC} compares the properties of the optical forbidden
lines of the HVC and LVC between epochs 1999 and 2010, based on 
[N {\sc ii}] $\lambda6583$ emission from the two epochs. The left panels compare spatial 
variations of their velocity centroids. The positions of the 1999 data points have been 
shifted to the 2010 epoch to account for their proper motions according to 
their respective velocity centroids. The velocity centroids of [Ne {\sc iii}] $\lambda3869$
were overlaid to indicate the regions based on the features of [Ne {\sc iii}] emission:
the brightest ``MicroJet'' from 0\farcs5 to 2\farcs5, non-detected ``InterKnot'' from
2\farcs8 to 5\farcs0, and ``Knot A$+$B'' from 5\arcsec\ to 8\arcsec. In the
right panels, we compare the integrated [N {\sc ii}] line profiles of the microjet between 1999 
and 2010. The 1999 line profile is obtained by summing over 0\farcs1 to 0\farcs7 
from the 1999 spectra,
corresponding to the innermost microjet. The 2010 line profiles are obtained in
two parts, one summing over the ``MicroJet'' region as the initial microjet in 2010,
and another summing over the ``InterKnot'' region.
For the HVC, the portion of the microjet in the 1999 data matches
the 2010 ``InterKnot'' region well in both velocity centroids and positions. This
suggests that the initial microjet moves out without much interaction with previous 
ejecta in this region. For the LVC, the emission is not spatially extended and no clear
interknot emission can be detected. The propagation or variation pattern is relatively
unclear and does not show clear trends between the two epochs.

      From comparisons of the spatial variations of velocity centroids, we
conclude that the emission from the innermost microjet is not stationary. 
The HVC centroids of [N {\sc ii}] $\lambda6583$ have changed
from a median value of $\sim -260$ \kmps\ in 1999 to $\sim -180$ \kmps\ in the 
2010 X-Shooter spectra. The shifted 1999 data points for the microjet
match those for the interknot region well in the 2010 X-Shooter data in both velocities
and positions. The transitions appear to be smooth, and the matches in kinematic
properties of between the two epochs suggest that the flow does not slow down 
significantly due to interactions with previous ejecta. The properties of both
[N {\sc ii}] and [Ne {\sc iii}] emission lines in the microjet are consistent with a recombination flow.
The [Ne {\sc iii}] $\lambda3869$ spatial profile is consistent with the flow recombining, 
and then being reionized, before the knot A$+$B. The average [S {\sc ii}] $\lambda\lambda6716/6731$ 
ratio increases from 0.5 to 0.8, corresponding to a decrease in $n_e$ from $\sim 10^4$ 
to $10^2$ cm$^{-3}$, at distances from 0\farcs5 to 4\farcs5, where the knot A$+$B starts 
to form with an increase in $n_e$ to $\sim 4000$ cm$^{-3}$. 
In the next subsection, we suggest a possible mechanism that might cause the
decrease in the HVC velocity in the microjet and momentarily account for the dominant
initial ionization in the microjet.

\subsection{Possible Origin of the HVC Variation -- Source of Flares in the System?}

Comparison of the line profiles at two different epochs shows an overall
decrease in jet speed from 1999 to 2010, without evidence for
deceleration by discrete shocks. One possible cause of such a decrease in 
jet speed is the increase in the launching radius of the jet. Assuming that the 
terminal speed of the jet is proportional to the Keplerian speed at the launching radius and 
that the proportion remains the same for the two configurations at different epochs, 
the ratio of the two velocities ($180/260 \approx 0.69$) corresponds to a change in the 
launching radius by a factor of $\sim 2$ ($(260/180)^2 \approx 2.1$). 
At the inclination angle to the DG Tau system, the deprojected flow velocity in 1999 is 
$\sim 330$ \kmps. As demonstrated in the X-wind sample fits to the jet and 
counterjet of RW Aur A in \cite{LS12}, the ratio of the terminal wind speed $v_{\rm w}$ 
to the Keplerian speed at the truncation radius, $v_{\rm x}$, can range from 1.5 to 3.5, 
depending on the mass loading in the magnetic field lines of the wind. 
If $v_{\rm w}/v_{\rm x} \sim 3$ is adopted, $v_{\rm x} \approx 110$ \kmps. The estimated stellar mass 
of DG Tau is $0.7\pm0.2M_\odot$ \citep[see][]{HEG95,Testi02}, and one may derive the 
truncation radius $R_{\rm x} = GM_\ast/v_{\rm x}^2 \approx 0.05\pm0.02$ au. 
In this scenario, the change in HVC centroids may
correspond to a change in truncation radius from 0.05 au in 1999 to 0.10 au in 2010.

     One possible explanation for the increase in the disk 
truncation radius is an expansion of the stellar magnetosphere, which 
pushes out the disk. At equilibrium, the inner edge of the disk can corrotate
with the star due to regulation by the magnetic field connecting the star and the disk.
When the magnetosphere expands, the inner edge of the disk will change 
from one equilibrium to a new equilibrium position further away from the star
with a lower Keplerian speed. During the transition, the angular speed of the
star and the inner disk edge may not coincide with each other, i.e., $\Omega_\ast
\neq \Omega_{\rm x}$. This may induce magnetic reconnections for the field lines
connecting the star to the disk, and a fraction of the energy release may contribute
to the hardening of X-rays. In the fluctuating X-wind theory \citep{SHGL},
the amount of X-ray luminosity generated through magnetic reconnection is proportional
to the square of the (dimensionless) dipole flux $\Phi_{\rm dx}$ of order unity and 
the energy scale for disk accretion, $\frac{GM_\ast \dot{M}_{\rm D}}{R_{\rm x}}$, 
both computed at the truncation radius. This proportional factor, which may range from
0.1 to 0.01, depends on the fraction of magnetic energy released, aside from that used 
for particle acceleration, and on the efficiency of energy conversion.
Taking the disk accretion rate 
$\dot{M}_{\rm D}$ to be of the same order as the mass accretion rate onto the star 
$\dot{M}_{\rm acc}$, the energy scale for disk accretion is of the order of $\sim 10^{33}$ 
erg\,s$^{-1}$. The resulting total X-ray luminosity would be of the order of
$\sim 10^{31}$ to $\sim 10^{32}$ erg\,s$^{-1}$ and is sufficient to power the hard X-ray source
close to DG Tau and to account for the ionization for [Ne {\sc iii}] $\lambda3869$
emission in the jet.

\section{SUMMARY}\label{summary}

      We have studied the kinematics and ionization properties of the DG Tau jet using 
archival {\it HST}/STIS spectra taken in 1999 January and new VLT/X-Shooter spectra
taken 11 years later. For both data sets, we identify an HVC
that is spatially extended and connected to outer knots and an LVC
that dominates within $\pm1\arcsec$ through Gaussian decomposition.
Optical [O {\sc i}], [S {\sc ii}], and [N {\sc ii}] line emission in the 1999 STIS spectra 
has various velocity centroids; [N {\sc ii}] has the fastest HVC at $\sim-260$ \kmps\
and LVC at $\sim-100$ \kmps\ and [S {\sc ii}] has the slowest HVC at $\sim-140$ \kmps\
and LVC at $\sim-60$ \kmps. In contrast, all the forbidden lines in the 2010 X-Shooter 
spectra show a relatively uniform HVC at $\sim-180$ \kmps\ and LVC at $\sim-70$ \kmps.

In the 2010 X-Shooter spectra, [Ne {\sc iii}] $\lambda3869$ is detected 
in the innermost microjet up to $\sim 3\arcsec$ and reappears as the unresolved knots A$+$B at 
$6\farcs5$, but is virtually undetected in knot C at $13\arcsec$. Both the HVC and LVC
of the microjet peak within $0\farcs4$ of the star. The HVC has a large line width
up to $\sim 200$ \kmps\ close to the source. The intensity of the HVC and LVC is comparable,
with a dereddened flux of $\sim 10^{-11}$ erg\,cm$^{-2}$\,s$^{-1}$, assuming $A_V\approx0.55$
\citep{Sch13L}. The profile at the knot A$+$B is singly peaked at $\sim-270$ \kmps, close to the 
[N {\sc ii}] HVC centroid in 1999. The flux at this knot is much fainter than the 
inner microjet by an order of magnitude. 

       We discuss possible origins of [Ne {\sc iii}] emission in light of the 
known X-ray sources in the DG Tau system. Shocks of a few hundred 
\kmps\ suffice to both collisionally ionize neon and heat the gas up to the megakelvin
temperatures seen in the soft X-ray source 0\farcs2 from DG Tau. However, the same 
shocks that account for the soft X-ray luminosity \citep{GML09} can produce only 
$\sim 10^{-2}$ of the [Ne {\sc ii}] flux observed. Soft X-ray irradiation 
from stationary stellar wind shocks \citep{GLS14} may partially contribute to
the ionization, but is swamped by the extant hard coronal X-ray source.
The hard coronal source, with impulsive flares up to 
$\sim 5\times10^{30}$ erg\,s$^{-1}$, may be able to account for the observed
[Ne {\sc iii}] flux, although recurring flares are needed to account for the 
extended emission. Possible soft X-ray or magnetic heating that maintains
the $10^5$ K gas associated with the C {\sc iv} jet \citep{Sch13L} may 
alternatively freeze the ionization in the jet and account for the observed spatial
extent of the [Ne {\sc iii}] microjet. 
The outer knot A$+$B may be reionized by a strong shock $v_{\rm shock} \gtrsim 100$ 
\kmps\ which would re-invigorate the [Ne {\sc iii}] emission and produce the extended
soft X-ray emission.

The 11 year baseline also allows us to examine changes in the velocity structure 
of the optical forbidden emission lines. The HVC velocity centroids decreased by about 30\% over 
these 11 years, while the LVC was little changed.
A possible explanation for the decrease in jet speed is an
increase in the truncation radius (as well as the jet launching radius) caused by
expansion of the stellar magnetosphere. The change in velocity would correspond
to an increase in radius by a factor of $\sim2$, assuming that the overall magnetospheric
configuration does not change after expansion. Magnetic
reconnection produced during the readjustment of the inner disk radius
may provide a sufficiently luminous X-ray source to explain the 
observed [Ne {\sc ii}] and  [Ne {\sc iii}] fluxes, but further X-ray and optical 
spectroscopic observations will be required to test this hypothesis.
 
\acknowledgements
The authors would like to thank the anonymous referee whose
constructive comments helped to improve the clarity of the manuscript.
This work was supported
by funds from the Academia Sinica Institute of Astronomy
and Astrophysics (ASIAA), and the Ministry of Science and Technology
(MoST) of Taiwan by grant NSC 102-2119-M-001-008-MY3.
This work is partly based on observations made with
ESO Telescopes at the La Silla Paranal Observatory under
programme 084.C-1095(A).
The {\it HST}/STIS spectra were obtained from the Mikulski
Archive for Space Telescopes (MAST) at the Space Telescope Science Institute
(STScI). STScI is operated by the Association of Universities
for Research in Astronomy, Inc., under NASA contract
NAS 5-26555.

\facility{{\it HST} (STIS), VLT:Kueyen (X-Shooter)}

\end{document}